\documentclass[12pt,preprint]{aastex}
\usepackage{color} 

\usepackage{epsfig}
\usepackage{subfigure}
\usepackage{graphicx}
\usepackage{amsmath}
\usepackage{natbib}
\usepackage{lscape}
\newcommand{\pa}{\partial}
\newcommand{\mb}{\boldsymbol}
\newcommand{\wt}{\widetilde}

\shorttitle{Particle-Gas Dynamics with Athena}
\shortauthors{Bai \& Stone}

\begin{document}


\title{Particle-Gas Dynamics with Athena: Method and Convergence}


\author{Xue-Ning Bai \& James M. Stone}
\affil{Department of Astrophysical Sciences, Princeton University,
Princeton, NJ, 08544} \email{xbai@astro.princeton.edu,
jstone@astro.princeton.edu}




\begin{abstract}
The Athena MHD code has been extended to integrates the motion of
particles coupled with the gas via aerodynamic drag, in order to study the
dynamics of gas and solids in protoplanetary disks and the formation of
planetesimals. Our particle-gas hybrid scheme is based on a second
order predictor-corrector method. Careful treatment of the momentum
feedback on the gas guarantees exact conservation. The hybrid scheme
is stable and convergent in most regimes relevant to protoplanetary disks.
We describe a semi-implicit integrator generalized from the leap-frog
approach. In the absence of drag force, it preserves the geometric
properties of a particle orbit. We also present a fully-implicit integrator
that is unconditionally stable for all regimes of particle-gas coupling.
Using our hybrid code, we study the numerical convergence of the
non-linear saturated state of the streaming instability. We find that gas
flow properties are well converged with modest grid resolution (128 cells
per pressure length $\eta r$ for dimensionless stopping time $\tau_s=0.1$),
and equal number of particles and grid cells. On the other hand, particle
clumping properties converge only at higher resolutions, and finer
resolution leads to stronger clumping before convergence is reached.
Finally, we find that measurement of particle transport properties resulted
from the streaming instability may be subject to error of about $\pm20\%$.
\end{abstract}



\section{Introduction}\label{sec:intro}

Aerodynamic coupling between gas and solid bodies plays a crucial role in the
dynamics of protoplanetary disks (PPDs) and in planetesimal formation (e.g.,
\citealp{Cuzzi_etal93,ChiangYoudin10}). In PPDs, the gaseous disk is partially
supported by the radial pressure gradient, and rotates at sub-Keplerian velocity.
On the other hand, the solid particles are not affected by the pressure gradient
and tend to orbit at Keplerian velocity, resulting in relative motion and gas drag.
The drag force is characterized by the stopping time $t_{\rm stop}$, where in the
absence of other forces, the particle velocity relative to gas would decrease with
time as $\exp{(-t/t_{\rm stop})}$.
It is most conveniently parameterized by the dimensionless stopping time,
$\tau_s=\Omega t_{\rm stop}$, where $\Omega$ is the Keplerian angular frequency.
$\tau_s$ measures the strength of coupling between gas and solids and depends
strongly on particle size and location in the PPDs. In the Epstein regime (for
particle size smaller than the gas mean free path, \citealp{Epstein24}), which is
most relevant in PPDs \citep{Weidenschilling77}, $\tau_s=\rho_sa\Omega/\rho_gc_s$,
where $\rho_s$ is the solid density of the particles, $a$ is the particle radius
(assuming spherical shape), $\rho_g$ is gas density and $c_s$ is gas sound speed.
For the standard minimum mass solar nebular (MMSN) model \citep{Hayashi81},
$\tau_s=1$ roughly corresponds to meter sized bodies at 1 AU.

In PPDs, dust grains settle to the disk midplane layer, and the interaction
between gas and solids may generate Kelvin-Helmholtz instability (KHI)
\citep{Weidenschilling80} and/or the streaming instability
\citep{YoudinGoodman05}. Hybrid simulations with both gas and solids are
necessary to uncover the non-linear dynamics in the dusty midplane layer.
In addition to the hydrodynamic solver, two ingredients are essential for such
hybrid simulations: First, large number of super-particles are
required to mimic the size distribution of real solids in the PPDs. Each
super-particle represents a swarm (billions or more) of real particles with
the same physical properties. Although a two-fluid approach can be used to
model the dusty disks (e.g., \citealp{GaraudLin04}), the particle approach is
necessary to account for a distribution of solid bodies with different physical
properties, and when the solids are largely collisionless. The second ingredient
is the aerodynamic interaction between gas and particles. In particular, the
momentum feedback from particles to gas must be included, which is essential for
the development of KHI and streaming instability.

Hybrid codes of the kind mentioned above have been developed by several
groups \citep{YoudinJohansen07,JohansenYoudin07,Balsara_etal09,Miniati10}.
One of the primary goals of this paper is to present the implementation of
hybrid particle-gas integration scheme into the Athena code, a new
grid-based code for compressible magnetohydrodynamics (MHD) based on higher
order Godunov methods. A comprehensive description of the implementation
and tests of the MHD algorithms are given in \citet{AthenaTech}. The underlying
hydrodynamic solver in Athena is similar to the methods used in \citet{Balsara_etal09}
and \citet{Miniati10}, but quite different from the finite difference methods,
used by \citet{YoudinJohansen07} and \citet{JohansenYoudin07} (hereafter, YJ07
and JY07 respectively). Our implementation of the hybrid
particle-gas scheme is different from any of the previous methods and has
several novel features. First of all, in solving the coupled equations of gas
and particles in hybrid simulations, the problem becomes difficult when the
particles are strongly coupled to the gas (i.e., $\tau_s\ll1$), so that the coupled
equations become very stiff. This is especially relevant to submillimeter
dust grains. We have developed a semi-implicit and a fully-implicit particle
integrators that can handle most particle-gas coupling regimes relevant
for PPDs. Secondly, our semi-implicit particle integrator naturally
generalizes from the leap-frog type integrator, and preserves geometric
properties of particle orbits. Thirdly, our hybrid scheme conserves linear momentum
to machine accuracy.

We show a suite of test problems to demonstrate the performance of our
particle integrators and the hybrid scheme. In particular, we show that using
the fully-implicit integrator is necessary when the particle stopping time is
less than the numerical time step. Further, we test and compare our code
with \citet{YoudinJohansen07}, \citet{JohansenYoudin07}, \citet{Balsara_etal09},
and \citet{Miniati10}. YJ07 provides two eigen-vectors of the unstable modes of
the streaming instability. Measuring
the linear growth rates of these unstable modes and comparing them with
theoretical values constitutes a stringent test problem of the hybrid code.
JY07 studied the non-linear saturation of the streaming instability with a set
of model parameters. They investigated a variety of physical quantities as
measured from their simulations. We show that our code reproduces these
published results.

Finally, we present a systematic study of the numerical convergence on
the simulations of the streaming instabilities.
Although JY07 performed some experiments to test the convergence of their
numerical results, they did not carry out a systematic study on this subject.
We pick two representative runs from JY07, and repeat the simulations with
different grid resolution, and different number of particles in the simulation
box. The results from our study provides useful insights in understanding the
uncertainties of various physical quantities as measured from such hybrid
simulations.

This paper is organized as follows. The formalism and the hybrid
particle-gas integration scheme of our code is presented in
\S\ref{sec:hybrid}. In \S\ref{sec:integrator}, we describe and discuss
two implicit particle integrators that solves the stiffness problem.
We provide a suite of test problems to demonstrate our code
performance in \S\ref{sec:test}. In \S\ref{sec:converge}, we
systematically study the numerical convergence of the streaming
instability in the saturation state. We summarize and conclude our
paper in \S\ref{sec:conclusion}.

\section[]{Hybrid Particle-Gas Scheme}\label{sec:hybrid}


\subsection[]{Formalism}\label{ssec:formalism}

Motivated by the study of the streaming instability in the context
of PPDs (e.g., \citealp{YoudinGoodman05}), we formulate the equations
with the local shearing sheet approximation
\citep{GoldreichLyndenBell65}. The source terms in this approximation
can be dropped to study other problems. We choose a local reference frame
located at a fiducial radius, corotating at the orbital angular velocity
$\Omega$. The dynamical equations are written using the Cartesian
coordinate, with $\hat{\mb{x}},\hat{\mb{y}},\hat{\mb{z}}$ denoting unit
vectors pointing to the radial, azimuthal and vertical direction. In this
non-inertial frame, the coupled equations of particles and gas read
\begin{equation}
\frac{d\mb{v}_i}{d
t}=2\mb{v}_i\times\mb{\Omega}+2q\Omega^2x\hat{\mb{x}}-\Omega^2z\hat{\mb{z}}
-\frac{\mb{v}_i-\mb{u}}{t_{\rm stop}}\ ;\label{eq:dustmotion}
\end{equation}
\begin{equation}\label{eq:gascont}
\frac{\pa\rho_g}{\pa t}+\nabla\cdot(\rho_g\mb{u})=0\ ,
\end{equation}
\begin{equation}
\frac{\pa\rho_g\mb{u}}{\pa t}+\nabla\cdot(\rho_g\mb{uu}+P_g\mb{I})
=\rho_g\bigg[2\mb{u}\times\mb{\Omega}+2q\Omega^2x\hat{\mb{x}}-\Omega^2z\hat{\mb{z}}+
\epsilon\frac{\overline{\mb{v}}-\mb{u}}{t_{\rm stop}}\bigg]\ .
\label{eq:gasmotion}
\end{equation}
In the above equations, $\rho_g$, $P_g$ denote the mass density and
pressure of the gas, $q\equiv d\ln\Omega/d\ln r$ is the background
shear parameter, with $q=3/2$ for Keplerian flow, and $\mb{u}$,
$\mb{v}$ denote velocities of the gas and particles in this reference
frame. The subscript ``$i$" in equation (\ref{eq:dustmotion}) represents
the $i$th particle. The particle stopping time due to gas drag,
$t_{\rm stop}$, depends on particle size and gas flow properties
\citep{Weidenschilling77}. Our code is capable of dealing with an
arbitrary number of different particle species (each particle species
has a different stopping time), but unless otherwise stated, we assume
single particle species with constant stopping time throughout this
paper for simplicity. In equation (\ref{eq:gasmotion}), $\overline{\mb{v}}$
stands for averaged particle velocity in the ``fluid element" (weighted by
mass), and $\epsilon$ denotes the local mass density ratio between
particle and gas $\epsilon=\rho_p/\rho_g$. This term represents
momentum feedback from the particles to the gas, written in the form
of treating particles as a fluid. The particle treatment of feedback
term is described in \S\ref{ssec:scheme}, and conservation of total
momentum is guaranteed. In this paper, we consider non-stratified disks
by neglecting vertical gravity terms in the equations above (i.e.,
the $\Omega^2z$ terms). We also neglect terms associated with the
magnetic field in this paper. They are handled by the underlying MHD
integrators in Athena \citep{AthenaTech}. An isothermal equation of state for the
gas is used throughout this paper, with $P=\rho_gc_s^2$. \footnote{
The drag between gas and solids dissipates energy and generates
heat. It is ignored in the isothermal gas.}

Our goal is to perform the local shearing box simulations \citep{HGB95},
where the radial boundary condition is periodic with additional
shear to account for differential rotation. Therefore, it is not
appropriate to include radial pressure gradient directly, which
is inconsistent with the periodic boundary conditions. Alternatively, one
can replace the pressure gradient by a constant radial force acting
on the gas ${\mb F}=2\eta v_K\Omega\hat{\mb x}$, pointing {\it outward}.
The quantity $\eta v_K$ measures the amount by which the gas (azimuthal)
velocity is reduced from the Keplerian value due to the radial pressure
gradient. In our code, instead, we find it more convenient not to modify
the hydrodynamic integrator, but to add a constant radial force on the
particles, pointing {\it inward}. Our treatment is mathematically
equivalent to the effect of a radial pressure gradient, but both
particle and gas (azimuthal) velocities are shifted to slightly
larger value, by $\eta v_K$.

An orbital advection scheme \citep{FARGO,Johnson_etal08} has been
implemented in Athena, which takes the advantage of the fact that
the above equations can be split into two systems, one of which
corresponds to linear advection operator with background flow
velocity $(q\Omega x)\hat{\mb{y}}$ and the other involves only
velocity fluctuations \citep{StoneGardiner10}. We use the orbital
advection scheme in all our simulations, not only because it is
it is faster, but also more accurate\footnote{This scheme is used
only in 3D shearing box simulations (e.g., \citealp{BaiStone10b}).
In this paper, where we perform 2D simulations in the
radial-vertical plane, the orbital advection scheme is not used.}.
The same technique can be implemented to the particles, with
equation (\ref{eq:dustmotion}) replaced by
\begin{equation}
\frac{d\mb{v}'_i}{d
t}=2(v'_{iy}-\eta v_K)\Omega\hat{\mb{x}}-(2-q)v'_{ix}\Omega\hat{\mb{y}}
-\Omega^2z\hat{\mb{z}}-\frac{\mb{v}'_i-\mb{u}'}{t_{\rm stop}}\
,\label{eq:dustFARGO}
\end{equation}
where $\mb{v}'=\mb{v}-(q\Omega x)\hat{\mb{y}}$, and
$\mb{u}'=\mb{u}-(q\Omega x)\hat{\mb{y}}$. An particle advection step
of is then carried out in parallel with the orbital advection of
gas. Also note that we've included the effect of gas radial
pressure gradient in the first term on the right hand side of
equation (\ref{eq:dustFARGO}).

\subsection[]{Predictor-Corrector Scheme}\label{ssec:scheme}

The MHD integrator in Athena adopts a directionally unsplit,
higher-order Godunov method, which conserves mass, momentum and
energy (when applicable) to machine accuracy. Our goal is to develop
a particle-gas hybrid integrator that is also conservative and at least
second order accurate. Two MHD integrators have been implemented in the
Athena code, including the cornered transport upwind (CTU) integrator
\citep{AthenaTech} and the van-Leer (VL) integrator \citep{StoneGardiner09}.
Our particle scheme is combined with the CTU integrator, which is more
accurate and less diffusive \citep{StoneGardiner09}. In our implementation,
the momentum (and energy) feedback to the gas is treated as source terms
(different from \citealp{Miniati10}), while the evaluation of transverse flux
gradient in the hydrodynamic solver is unchanged. The gas continuity
equation is automatically handled in the Godunov scheme, with no
modifications to the code needed. The hybrid particle-gas scheme adopt a
predictor-corrector approach, which is described below.

To demonstrate the numerical algorithm, we rewrite the coupled particle-gas
momentum equations. For the gas momentum equation, we simplify
the left hand side of equation (\ref{eq:gasmotion}) to a Lagrangian
derivative, since we do not modify the calculation of the flux gradients
in the CTU integrator. We use ${\mb f}({\mb x},{\mb v})$ and
${\mb g}({\mb x},{\mb u})$ to denote source terms for the acceleration of
particles and gas respectively, due to forces other than the drag force
(e.g., Coriolis force, tidal force and the global pressure gradient). The
formalism in \S\ref{ssec:formalism} can be summarized as
\begin{equation}
\frac{d\mb{v}_i}{dt}={\mb f}_i-\frac{\mb{v}_i-\mb{u}}{t_{\rm stop}}\ ;\label{eq:dustmotion1}
\end{equation}
\begin{equation}
\frac{d(\rho_g\mb{u})}{dt}=\rho_g{\mb g}+\rho_p\frac{\overline{\mb{v}}-\mb{u}}{t_{\rm stop}}\ .
\label{eq:gasmotion1}
\end{equation}
We do not distinguish between (${\mb u}, {\mb v}$) and (${\mb u}', {\mb v}'$)
here because it does not affect our description of the numerical algorithms.
Our predictor-corrector scheme, which integrates the coupled equations
from time step $t^{(n)}$ to $t^{(n+1)}$, can be illustrated as
\begin{subequations}\label{eq:scheme}
\begin{align}
(\rho_g\mb{u})^{(n+1/2)}&=(\rho_g\mb{u})^{(n)}+\frac{h}{2}(\rho_g{\mb g})^{(n)}
+\sum_iW({\mb x}-{\mb x}_i^{(n)})\Delta{\mb p}_i^{\rm pred}\ ,\label{eq:scheme1}\\
\mb{v}_i^{(n+1)}&=\mb{v}_i^{(n)}+h[{\mb f}({\mb v}_i, {\mb u}^{(n+1/2)}({\mb x}_i))
-(\mb{v}_i-{\mb u}^{(n+1/2)}({\mb x}_i))/t_{\rm stop}]\ ,\label{eq:scheme2}\\
(\rho_g\mb{u})^{(n+1)}&=(\rho_g\mb{u})^{(n)}+h(\rho_g{\mb g})^{(n+1/2)}+
\sum_iW({\mb x}-{\mb x}_i^{(n+1/2)})\Delta{\mb p}_i^{\rm corr}\ ,\label{eq:scheme3}
\end{align}
\end{subequations}
In the above, (\ref{eq:scheme1}) and (\ref{eq:scheme3}) are generalizations
of the predictor and corrector step of the MHD integrator, and particle
feedback to gas is expressed as $\Delta{\mb p}_i^{\rm pred}$ and$\Delta{\mb p}_i^{\rm corr}$
for the two steps respectively. Their expressions are given in the following
paragraphs. $W$ is the weight function for interpolation (see next paragraph).
(\ref{eq:scheme2}) represents the particle integrator, which we will discuss
in detail in \S\ref{sec:integrator}. Note that in the bracket on the right hand
side of this equation, particle quantities ${\mb x}_i$ and ${\mb v}_i$ can be
combinations of step $(n)$ and step $(n+1)$ quantities, depending on the particle
integrator.

The calculation of the drag force experienced by particles requires
interpolation of grid quantities to the particle location on the
grid. The interpolation scheme is described by the weight function
$W({\mb x}-{\mb x}_i)$. For consistency, the same interpolation scheme
is used to distribute the feedback from individual particles to the gas
grid points, as in equation (\ref{eq:scheme}). In order avoid spurious
numerical artifacts in the hybrid scheme, the weight function $W$ should
satisfy certain constrains. It should be continuous over the
computational domain to avoid sharp transitions, and $W(\Delta{\mb x})$
should be non-negative for any $\Delta{\mb x}$ to reduce noise
\citep{YoudinJohansen07}. The interpolation should be accurate enough to
minimize errors. In particular, interpolation error should not be much
worse than the error from the spacial reconstruction of the MHD integrator
(we use the third order piecewise parabolic method). Finally, interpolation
is time consuming, so the scheme should be as efficient as possible. We
have compared three interpolation schemes (cf.\citealp{BirdsallLangdon05},
YJ07), namely, cloud-in-a-cell (CIC), triangular-shaped cloud (TSC) and
quadratic polynomial (QP). The CIC scheme is simple but inaccurate and
noisy, the QP scheme is the most accurate but not continuous. Similar to
JY07, we choose the TSC interpolation scheme throughout this paper.


In the predictor step, the momentum feedback from individual particles
is calculated from direct force evaluation, multiplied by half a time
step. However, when the particle stopping time is small, the drag force
diverges, and the error in the velocity calculation and interpolation
can be substantially amplified. We note that in the absence of other
forces, the particle velocity would approach gas velocity as
$\exp{(-t/t_{\rm stop})}$. Therefore, we modify the predictor step
momentum feedback (from individual particle ``$i$") into
\begin{equation}\label{eq:fbpred}
\Delta{\mb p}_i^{\rm pred}=\frac{m_i({\mb v}_i-{\mb u})}
{\max{(t_{\rm stop}},h)}\frac{h}{2}\ ,
\end{equation}
where $m_i$ is the particle mass. We do not use the exponential
expression because it is derived by assuming drag force only. Our
treatment is the same order accurate as the exponential expression
(first order) when $t_{\rm stop}\leq h$, and when $t_{\rm stop}\geq h$,
it ensures exact force balance at equilibrium state\footnote{For
example, when testing the linear growth rate of the streaming
instability (in \S\ref{sec:test}), one starts from the NSH equilibrium.
Our treatment allows NSH equilibrium to be satisfied exactly.}. The
individual particle feedback is then distributed to neighboring grid
cells. As a source term, the momentum feedback is divided by gas
density and is added to the left and right states of primitive
variables (gas velocity) between step 1 and step 2 of the CTU
algorithm as described in \citet{AthenaTech}.

For the feedback calculation in the corrector step, note that the
particles have already evolved for a full time step, we can
calculate the momentum difference of individual particles between
the two steps $(n)$ and $(n+1)$. Let ${\mb F}_c$ denote forces
experienced by particles other than the drag force. Since
${\mb F}_c$ is generally non-stiff (e.g., Coriolis force), we can
obtain momentum feedback from particle $i$ to be
\begin{equation}\label{eq:fbcorr}
\Delta{\mb p}_i^{\rm corr}={\mb p}_i^{(n+1)}-{\mb p}_i^{(n)}
-{\mb F}_{c,i}^{(n+1/2)}h\ ,
\end{equation}
where ${\mb F}_{c,i}^{(n+1/2)}$ is evaluated at ${\mb x}^{(n+1/2)}$,
the midpoint between ${\mb x}^{(n)}$ and ${\mb x}^{(n+1)}$, which
ensures second order accuracy. This treatment is conservative and
guarantees exact momentum conservation. Moreover, it avoids the
potential stiffness due to the direct evaluation of the drag force. To
distribute the feedback to grid points, we again take the force
location at ${\mb x}^{(n+1/2)}$, as indicated in equation
(\ref{eq:scheme3}). The corrector step feedback is added to the end of
the gas integrator.

If one considers the disk thermodynamics (not treated in this paper),
the heat generated by friction needs to be deposited to the energy
equation of the gas, with energy dissipation rate (per unit volume)
\begin{equation}
\dot{\mathcal E}=\rho_p(\overline{\mb v}-{\mb u})^2/t_{\rm stop}\ .
\end{equation}
The heating term is contributed from the work done by the pressure
gradient and is typically only a small amount compared with the disk
thermal energy budget (see YJ07 for more details). Numerically,
potential stiffness problem also exists since $t_{\rm stop}$ is
present in the denominator. In practice we rewrite the energy deposition rate
from particle $i$ to be $\dot{\mathcal E}_i=(\Delta p_i)^2t_{\rm stop}/m_ih^2$,
where $m$ is particle mass and $\Delta p_i$ is taken from equations
(\ref{eq:fbpred}) and (\ref{eq:fbcorr}) for predictor and corrector steps.

The overall accuracy of our hybrid scheme is second order, less than
the Pencil code, which is a finite-difference code with higher order
accuracy in smooth flows. However, our code is fully conservative, and
it does not need to be stablized by artificial hyper-viscosity. It also allows the
development of implicit particle integrator (see next section) much easier.
Recently,  \citet{Balsara_etal09} has proposed a similar predictor-corrector
hybrid scheme for particle-gas dynamics, but approximation is made
in the corrector step (see their equation (16) and the discussion that follows)
and is less accurate when dust dominates local density. \citet{Miniati10} has
described another hybrid scheme that is fully implicit, but assumes only one
particle species. Our predictor-corrector scheme is simpler than these other
approaches, allows an arbitrary number of particle species,
while we will show in \S\ref{sec:test} and \S\ref{sec:converge} that our code
performance is at least comparable to all these codes.

One disadvantage of our predictor-corrector hybrid scheme is that it is
intrinsically explicit, and may cause numerical instability when the
coupled equations are stiff. Here, the stiffness is caused by the
parameter $\epsilon$, the {\it local} particle to gas mass ratio. The design
of the particle integrators assumes that particles move in the gas
velocity field. In the regime where $t_{\rm stop}\lesssim h$, our
semi-implicit and fully-implicit schemes force particles to be strictly
coupled to the gas. However, when $\epsilon\gg1$, gas is expected to
follow the particles. This situation may result in unphysically large growth
of particle and gas velocities, making the hybrid code unstable. More
specifically, we define the stiffness parameter
\begin{equation}
\chi\equiv\sum_k\epsilon_k h/\max{(t_{{\rm stop},k},h)}\ .\label{eq:stiffness}
\end{equation}
In the above expression we have generalized our analysis to include a
size distribution of particles, and subscript ``$k$" labels different
particle types. Our hybrid scheme can become unstable when $\chi$
exceeds order unity. Our experiments show that the threshold value of
$\chi$ is about $3-5$ (depending on the actual problem).

One way to remedy the stiff particle mass loading problem discussed
above is to make the overall hybrid scheme implicit\footnote{Another
approach is to artificially reduce the momentum feedback by
$\max{[\chi,1]}$ in the predictor step (not in the corrector step,
otherwise momentum conservation would break down). This approach
reduces the accuracy of the code to first order, but improves the
stability.}. This is relatively easy to do if all particles have a
single size, and the treatment will be similar to a two-fluid
approach \citep{Miniati10}. However, the two-fluid treatment is not
easily generalized to a size distribution of particles. This is
because particles with different sizes are (indirectly) coupled to
each other via interactions with the gas, and evolving the coupled
equations implicitly requires solving an inverse matrix of rank
$N+1$ at each grid cell, where $N$ is the number of particle size
bins. The matrix is non-diagonal due to the coupling \footnote{An
example of such a matrix is given in the Appendix A of
\citet{BaiStone10b}.}. As $N$ gets larger and larger, the algorithm
becomes more and more complicated and computationally prohibitive.

Alternatively, in regions where $\chi$ exceeds certain thresholds,
one can effectively increase the value of $t_{\rm stop}$ for all local
particles to bring down the value of $\chi$, which guarantees stability,
as is adopted in the Pencil Code (A. Johansen, private communication,
2010). Physically, this means in dense particle clumps, particles can
move more freely and are less affected by gas drag, while just
sufficient momentum feedback is added to the gas for it to follow the
motion of particles without  causing any numerical instability. We
have implemented this technique that enforces $\chi<3$ in our code,
although in practice, this feature is turned off due to the reasons below.

Fortunately, in the context of PPDs, the overall dust to gas mass
ratio is about one percent. Therefore, in an average sense,
$\overline\chi$ is much less than unity. $\overline\chi$ can become
larger when dust grains settle towards the disk midplane. The typical
time step in the simulations is about $\Omega h=10^{-4}-10^{-3}$. For
dust grains with stopping time $\tau_s\lesssim10^{-3}$, one can easily
show that even very weak turbulence is able to make these solids
suspended in the disks, keeping $\overline\chi$ relatively small at
disk midplane\footnote{With the standard $\alpha$ prescription
\citep{ShakuraSunyaev73}, the average value of $\chi$ in disk midplane
can be estimated by $\overline{\chi}\sim Z(\max{[\Omega h,\tau_s]}/\alpha)^{1/2}$,
where $Z$ is the ratio of dust surface density over gas surface density,
typically $Z=0.01$. Thus fairly weak turbulence $\alpha=10^{-5}$ is
sufficient to keep $\overline{\chi}$ well below unity.}. Concentration
of particles can raise $\chi$ locally. In the case of the streaming
instability, however, particle concentration is most efficient for
marginally coupled solids $\tau_s\sim1$. Clumping of these particles
only raises $\chi$ very slowly according to equation
(\ref{eq:stiffness}). In our simulations of the SI in \S\ref{sec:converge}, 
the maximum value of $\chi$ never exceed the threshold $\chi=3$ (see
\S\ref{ssec:concentration} for more details). We emphasize that the
value of $\chi$ only determines stability, but does not constrain the
maximum particle density. In some of our stratified disk simulations
with large solid abundance (5-7 times super-solar, \citealp{BaiStone10c}),
we do observe the numerical instability and have to reduce the time step
in our calculations.

\section[]{Particle Integrator}\label{sec:integrator}

In this section, we describe in detail the particle integrator that
have been implemented in our particle-gas hybrid scheme [i.e.,
equation (\ref{eq:scheme2})]. The overall problem for the particle
integrator is
\begin{equation}
\frac{d{\mb x}}{dt}={\mb v}\ ,\qquad \frac{d{\mb v}}{dt}={\mb
a}[{\mb v}, {\mb x}, {\mb u}^{(n+1/2)}({\mb x})]\ ,
\end{equation}
where ${\mb a}$ denotes the acceleration due to all the forces,
including the gas drag, and following the convention of equation
(\ref{eq:scheme}). As suggested in equation (\ref{eq:scheme2}),
we use half time step gas velocities ${\mb u}^{(n+1/2)}$ to avoid
tracking the evolution of gas, which also ensures second order
accuracy. For the sake of simplicity, in the remaining of this
section, we will drop the superscript $^{(n+1/2)}$ in the gas
quantities.

As we have noted before, the gas drag term becomes stiff for
strongly coupled particles.
We have developed two second-order implicit particle integrators.
Each integrator has its own pros and cons. In our code, we allow
different species of particles to be pushed by different integrators,
which enable us to integrate particles with any stopping time while
maintaining the geometric properties of particle orbits.

\subsection[]{A Semi-implicit Integrator}\label{ssec:semi-imp}

Our first integrator is a semi-implicit scheme based on the
Crank-Nicholson method. The basic algorithm is the following.

\begin{equation}
\begin{split}
{\mb x}'=&{\mb x}^{(n)}+h{\mb v}^{(n)}/2\ ,\\
{\mb v}^{(n+1)}=&{\mb v}^{(n)}+h{\mb a}[({\mb v}^{(n)}+{\mb
v}^{(n+1)})/2, {\mb x}', {\mb u}({\mb x}')]\ ,\\
{\mb x}^{(n+1)}=&{\mb x}'+h{\mb v}^{(n+1)}/2\ ,\\
\end{split}\label{eq:semiimp}
\end{equation}
where ${\mb x}'={\mb x}^{(n)}+h{\mb v}^{(n)}/2$ is the predicted
particle position at $t^{(n+1/2)}$, and is used to evaluate the
stopping time and the gas velocity. This scheme is semi-implicit in
the sense that the velocity update depends on velocities in both
step $(n)$ and step $(n+1)$. Converting to explicit form, we find
\begin{equation}
{\mb v}^{(n+1)}-{\mb v}^{(n)}=h\Lambda^{-1}{\mb a}({\mb v}^{(n)},
{\mb x}', {\mb u}({\mb x}'))\ ,\qquad
\Lambda=1-\frac{h}{2}\frac{\pa\mb a}{\pa \mb v}\ ,\\
\end{equation}
where the Jacobian $\pa{\mb a}/\pa{\mb v}$ is evaluated at ${\mb
x}'$, with
\begin{equation}
\frac{\pa\mb a}{\pa \mb v}=\begin{pmatrix}
-1/t_{\rm stop} & 2\Omega & 0\\
-(2-q)\Omega & -1/t_{\rm stop} & 0\\
0 & 0 & -1/t_{\rm stop}\\
\end{pmatrix}\ .
\end{equation}
We always use the orbital advection algorithm \citep{StoneGardiner10},
therefore the $(2-q)$ factor in the matrix element. After calling the
particle integrator, we need to shift the particle positions by an
amount $-q\Omega xh\hat{\mb{y}}$. In practice, we replace $x$
by $[x^{(n)}+x^{(n+1)}]/2$. We emphasize that using the particle
advection scheme is important for improving the accuracy of
the particle integrator, and especially, preserve the geometric
properties of particle orbits (see below).

This semi-implicit integrator has close analogy to the Boris
integrator due to the similarity between Coriolis force and Lorentz
force. It is essentially a Leap-Frog integrator, in the form of
``Drift-Kick-Drift" (DKD). In \S\ref{sec:test}, we will see that
in the limit $t_{\rm stop}=\infty$, this integrator preserves
geometric orbital properties exactly. Recently, \citet{Quinn_etal10}
has proposed a symplectic particle integrator for Hill's equations,
which is essentially in the form of ``Kick-Drift-Kick". In their
work, particle advection was not implemented, making their formula
slightly more complicated than ours.

This integrator, although is implicit, can still cause problems when
the particle stopping time approaches zero (i.e., when $t_{\rm
stop}\ll h$). In this limit, the position update reduces to ${\mb
x}^{(n+1)}={\mb x}^{(n)} +h{\mb u}({\mb x}')$, which is indeed
second order accurate. However, the velocity update
reduces to ${\mb v}^{(n)}+{\mb v}^{(n+1)}=2{\mb u}({\mb x}')$. If
the initial velocity difference between particle and gas is large,
then particle velocity will oscillate around the gas velocity
without damping. More seriously, the evolution of gas velocity may
even amplify the velocity difference between particle and gas,
causing a runaway.
Our experiences indicate that this integrator works safely at least for
$t\gtrsim0.2h$. One can construct other similar second order
semi-implicit, Crank-Nicholson type schemes. Nevertheless, the only
other possibility one can achieve in the $t_{\rm stop}\rightarrow0$ limit is
${\mb v}^{(n)}+{\mb v}^{(n+1)}={\mb u}({\mb x}^{(n)})+{\mb u}({\mb x}^{(n+1)})$,
which is prone to the same problem. Also, other schemes no longer
maintain geometric properties of particle orbits.

\subsection[]{A Fully-implicit Integrator}\label{ssec:full-imp}

The apparent weakness of any types of the semi-implicit method leads
us to develop an absolutely stable, second order integration scheme. To
be absolutely stable, we demand the method to be {\it fully implicit},
that is, the velocity update to step $(n+1)$ depends only on velocity at
step $(n+1)$. For a simple ordinary differential equation $dy/dt=f(y)$,
a second order fully-implicit scheme can be constructed as
\begin{equation}
y^{(n+1)}=y^{(n)}+hf(y^{(n+1)})/2+hf[y^{(n+1)}-hf(y^{(n+1)})]/2\ .
\end{equation}
Other second order fully-implicit schemes are possible, but we will
stick to this specific form. The implementation of this method to the
particle integrator is illustrated as follows
\begin{equation}
\begin{split}
{\mb x}'&={\mb x}^{(n)}+h{\mb v}^{(n)}\ ,\\
{\mb v}^{(n+1)}&={\mb v}^{(n)}+\frac{h}{2}{\mb a}\bigg({\mb
v}^{(n+1)}, {\mb x}'\bigg)+\frac{h}{2}{\mb a}\bigg({\mb
v}^{(n+1)}-h{\mb a}[{\mb v}^{(n+1)}, {\mb x}')], {\mb x}^{(n)}\bigg)\ ,\\
{\mb x}^{(n+1)}&={\mb x}^{(n)}+\frac{h}{2}\bigg({\mb v}^{(n)}+{\mb v}^{(n+1)}\bigg)\ ,\\
\end{split}\label{eq:fullimp}
\end{equation}
where we have omitted the fluid argument ${\mb u}^{(n+1/2)}$ for
conciseness, and we assume that ${\mb u}$ is evaluated at the same
position (${\mb x}'$ or ${\mb x}^{(n)}$) as in ${\mb a}$. It can be
easily shown that in the limit $t_{\rm stop}=0$, the above equations
demand that ${\mb v}^{(n+1)}={\mb u}({\mb x}')$. Although this is not
second order accurate, the position update is indeed second order.
Therefore, we achieve second order accuracy with absolute stability.

The main work is to update the velocity, and to the second order, we
find
\begin{equation}
\begin{split}
{\mb v}^{(n+1)}-{\mb v}^{(n)}&=\frac{h}{2}\Lambda^{-1}\bigg[{\mb
a}({\mb v}^{(n)},{\mb x}^{(n)})+\bigg(1-h\frac{\pa\mb a}{\pa \mb
v}\bigg|_0\bigg){\mb a}({\mb v}^{(n)},{\mb x}')\bigg]\ ,\\
\Lambda&=1-\frac{h}{2}\bigg(\frac{\pa\mb a}{\pa \mb
v}\bigg|_1+\frac{\pa\mb a}{\pa \mb v}\bigg|_0-h\frac{\pa\mb a}{\pa
\mb v}\bigg|_0\frac{\pa\mb a}{\pa \mb v}\bigg|_1\bigg)\ ,\\
\end{split}
\end{equation}
where the subscript ``0" means to evaluate the Jacobian at ${\mb
x}^{(n)}$, ``1" means to evaluate at ${\mb x}'$ (note that the
stopping time can depend on position in the most general case).
The inverse matrix $\Lambda^{-1}$ can be evaluated analytically
without any trouble, since it involves only the inversion of a
$2\times2$ matrix. Alternatively, it suffices to evaluate the
inverse matrix up to first order in $h$, except for terms
containing $1/t_{\rm stop}$, where all orders of $h$ should be
kept. The result is
\begin{equation}
\Lambda^{-1}=(1+b)^{-2}\begin{pmatrix}
1+b & 2h\Omega & 0\\
-(2-q)h\Omega & 1+b & 0\\
0 & 0 & 1+b\\
\end{pmatrix}\ ,\quad
b\equiv\frac{1}{2}\bigg(\frac{h}{t_{\rm stop0}}+\frac{h}{t_{\rm
stop1}}+\frac{h^2}{t_{\rm stop0}t_{\rm stop1}}\bigg)\ .\label{eq:inverse}
\end{equation}

From equation (\ref{eq:inverse}), we see that as $t_{\rm stop}\rightarrow0$,
$\Lambda^{-1}\rightarrow b^{-1}\sim 2t_{\rm stop}^2/h^2\rightarrow0$.
Therefore, this integrator is stable for any $t_{\rm stop}>0$. One
disadvantage of this integrator is that it no longer preserves geometric
properties of particle orbits as $t_{\rm stop}\rightarrow\infty$. In fact, like
the explicit method, any fully-implicit method always fail to preserve
such geometric properties (see \S\ref{ssec:epicycle}). Therefore, we
will in most cases use the semi-implicit particle integrator, while we
switch to this fully-implicit integrator for particles with
$t_{\rm stop}\lesssim h$.

An alternative way of integrating the strongly coupled particles is to
use the ``short friction time" approximation introduced by
\citet{JohansenKlahr05}. In this approximation, particles can be
considered as being carried with the gas, while maintaining a small
drift velocity due to some external force ${\mb F}_p$. Meanwhile,
the gas also feels some external force ${\mb F}_g$ other than the drag.
${\mb F}_g$ is in general different from ${\mb F}_p$, since particles
don't have pressure support, and generally don't feel the magnetic field.
In this approximation, particle velocity is assumed to be always equal
to the termination velocity
\begin{equation}\label{eq:terminal}
{\mb v}_{\rm term}({\mb x})\approx {\mb u}({\mb x})+t_{\rm stop}
({\mb F}_p-{\mb F}_g)\ .
\end{equation}
One can then easily integrate particle orbit based on this expression
to any order of accuracy. We note that leading truncation error of this
approximation is of the order $\max[(t_{\rm stop}|\nabla{\mb u}|)^2,$
$t_{\rm stop}^2|\nabla{\mb F}|, t_{\rm stop}^2\Omega|\nabla{\mb u}|]$,
where ${\mb F}={\mb F}_p-{\mb F}_g$. Roughly speaking, the short
friction time approximation is applicable for stopping time
$t_{\rm stop}\ll\min(\Omega^{-1}, |\nabla{\mb u}|^{-1})$. Converting
the above into a second order integrator further introduces truncation
error of the order $(\Omega h)^2$. Therefore, this integrator may
perform equally well as our fully-implicit integrator only when
$t_{\rm stop}<h$. The original implementation of
\citet{JohansenKlahr05} did not include the momentum feedback,
thus can not be used to study the SI. However, it can be extended to
include feedback as described in \S\ref{ssec:scheme}. Nonetheless,
we do not implement this integrator in our code since we have the
fully-implicit integrator at hand.

\section[]{Code Tests}\label{sec:test}


\subsection[]{Epicycle Test}\label{ssec:epicycle}

We begin by examining the performance of the particle integrators
in the weak coupling limit, i.e., $t_{\rm stop}=\infty$, so gas drag
does not enter the problem. From the particle equation of motion
(\ref{eq:dustmotion}), the particle trajectory follows an epicyclic
orbit:
\begin{equation}
x(t)=A\cos\omega t\ , \qquad y(t)=-\frac{2\Omega}{\omega}A\sin\omega t\ ,
\qquad \omega=\sqrt{2(2-q)}\Omega\ .
\end{equation}
where $x(t), y(t)$ denote radial and azimuthal directions relative
to domain center, and $A$ is the radial amplitude of the oscillation.
Epicyclic motion conserves total energy
\begin{equation}
E=\frac{1}{2}(\dot{x}^2+\dot{y}^2)-q\Omega^2x^2=(2-q)\Omega^2A^2\ .
\end{equation}

\begin{figure*}
    \centering
    \includegraphics[width=150mm,height=60mm]{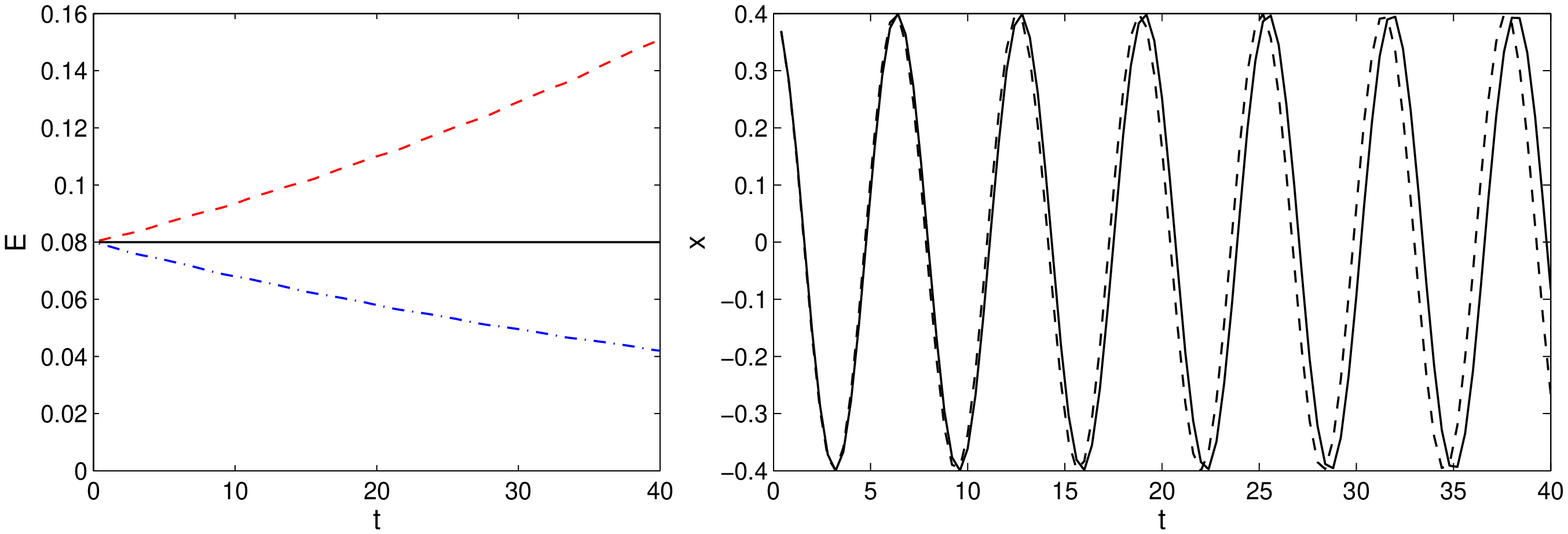}
  \caption{Results of the epicycle test, using $\Omega=1$, $q=3/2$, and
  one test particle with initial amplitude $A=0.4$, at fixed time step
  $h=0.4$. Left: Total energy of the test particle integrated using explicit
  (dashed), semi-implicit (solid) and fully-implicit (dash-dotted) methods.
  Right: Particle radial orbit integrated using the semi-implicit method
  (solid) compared with analytical solution (dashed).}\label{fig:epicycle}
\end{figure*}

We integrate a test particle and follow its epicyclic orbit using
different particle integrators and examine the particle trajectory
and energy conservation. In particular, we consider three 2nd order
integrators, namely, the semi-implicit and fully-implicit
integrators introduced in \S\ref{sec:integrator}, and we add an
explicit integrator based on the modified Euler method for
comparison purpose. In Figure \ref{fig:epicycle}, we see that the
semi-implicit method conserves total energy exactly\footnote{This
result is valid only when the orbital advection algorithm is used.
Otherwise, similar to the leap-frog method, the numerical Hamiltonian
is time-dependent and energy is not conserved, but oscillates around
the true value (if the time step is not constant, the energy may even
gradually deviate from the true value).}.
The particle orbit is closed, and the truncation error exhibits as a
phase shift relative to the analytical solution. The phase error
diminishes as $h^2$ (not shown in the figure), as expected. Note that
in order to make the errors significant, we have chosen a rather large
time step $h=0.4\Omega^{-1}$. In typical numerical simulations, the
time step and thus phase error is much smaller. From Figure
\ref{fig:epicycle}, we also see that the test particle gains energy
and drifts away from the guiding center when using the explicit method,
while it loses energy and gyrates into the guiding center when using
the fully-implicit method. This result is quite general.

This test demonstrates that the semi-implicit method preserves the
geometric properties of particle orbits, and moreover, it is efficient
because it evaluates the drag force (therefore interpolation) only once
per step. Therefore, in most applications we prefer to use this
integrator.

\subsection[]{Particle-Gas Deceleration Test}

\begin{figure}
    \centering
    \includegraphics[width=75mm,height=70mm]{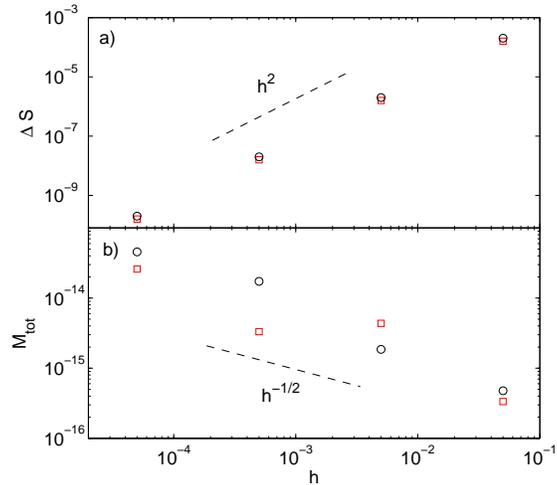}
  \caption{Error in position (a) and total momentum density (b) as a
  function of time step length $h$ in the particle-gas deceleration test. We
  choose the particle to gas mass ratio to be 1, and the initial velocities
  of particles and gas are both 1, but in opposite directions. The particle
  stopping time is set to $t_{\rm stop}=2$. Errors are measured at $t=1$.
  Red squares: semi-implicit integrator; Black circles: fully-implicit
  integrator.}\label{fig:collision}
\end{figure}

In the second problem, we consider the motion of a collection of particles
in a uniform gas. The spatial distribution of the particles is uniform. Gas
drag and feedback are included in the test. We work in the frame where the
total momentum in particles and gas is zero. Let ${\mb w}_0$ be the
particle initial velocity, then all the particles evolve as
\begin{equation}
S(t)=\frac{w_0t_{\rm stop}}{1+\epsilon}[1-e^{-(1+\epsilon)t/t_{\rm stop}}]\ ,
\end{equation}
where $\epsilon$ is the overall particle to gas mass ratio, and $S$ is
the distance a particle has traveled.

In our first test, we choose $\epsilon=1, t_{\rm stop}=2, w_0=1$, and
gas with density $\rho_g=1$. We evolve the system to $t_e=1$
with constant time step $h$ using the semi-implicit and the
fully-implicit integrators. In Figure \ref{fig:collision}a we plot the
position error $\Delta S$ at $t_e=1$ as a function of $h$. It is clear
that the cumulative position error scales with $h^2$, indicating that
our particle-gas hybrid code has achieved second order accuracy.
Figure \ref{fig:collision}b shows the total momentum density at
$t_e=1$ as a function of $h$. We see that our hybrid scheme conserves
linear momentum within machine accuracy (to the level of $10^{-16}$).
The total momentum at $t_e=1$ increases with decreasing $h$, which
reflects the accumulation of round-off error with increasing number of
time steps. For uncorrelated round-off errors, one would expect
$h^{-1/2}$ scaling. The slightly steeper slope reflects a certain level
of correlation in the round-off errors.

\begin{figure}
    \centering
    \includegraphics[width=75mm,height=75mm]{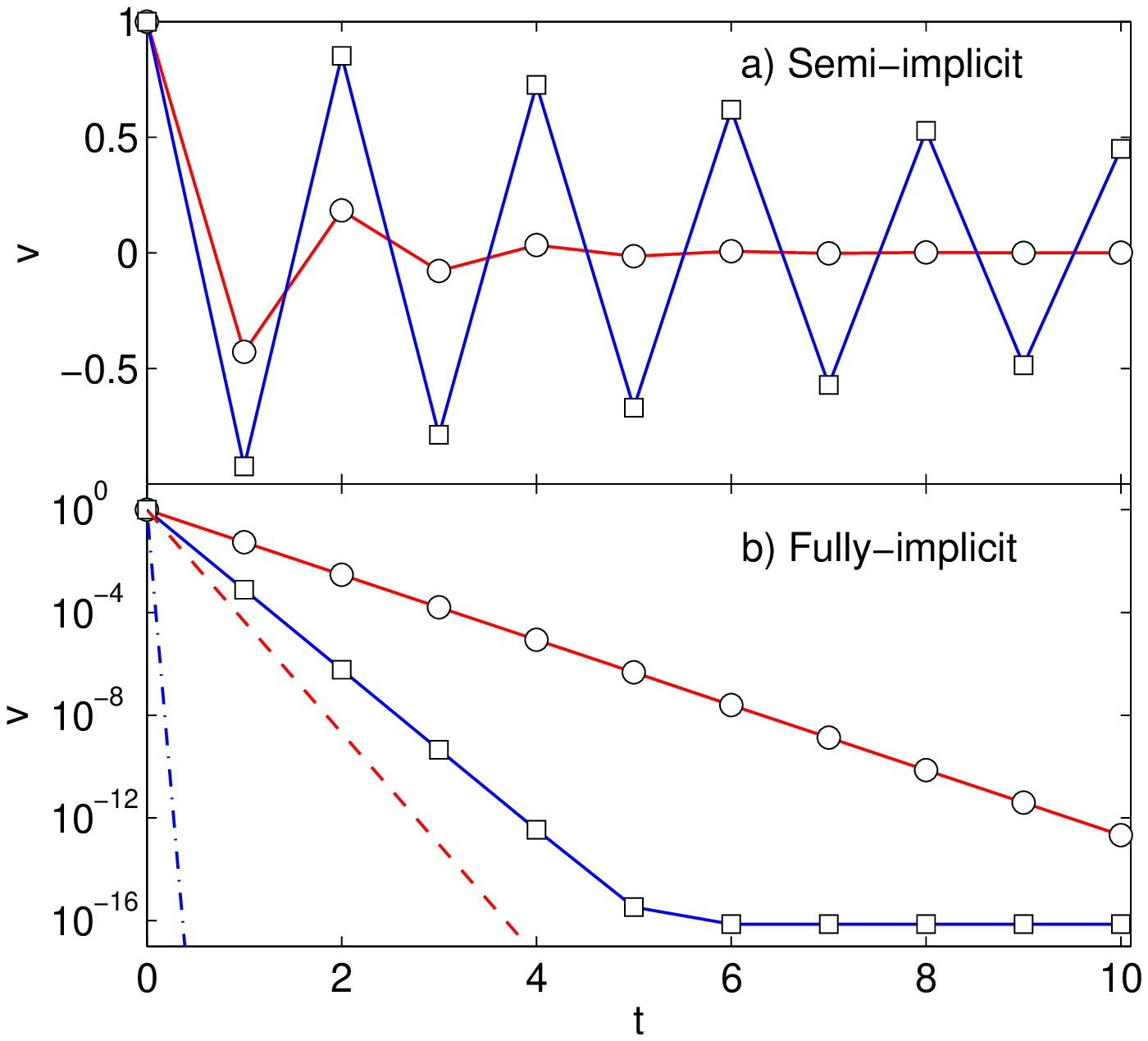}
  \caption{Numerical evolution of particle velocity in the stiff regime
  of the particle-gas deceleration test. The time step is fixed to $h=1$. Panels
  (a) and (b) show the results from the semi-implicit integrator and
  fully-implicit integrator respectively. Black circles connected by
  red solid lines: $t_{\rm stop}=0.2$. Black squares connected by blue
  solid lines: $t_{\rm stop}=0.02$. In panel (b), we further show
  expected evolution of velocity in the two cases: Red dashed line for
  $t_{\rm stop}=0.2$ and blue dash-dotted line for $t_{\rm stop}=0.02$.
  Note that in panel (a) we use a linear scale while in panel (b) we use
  a log scale.}\label{fig:collision2}
\end{figure}

Next we test the integrators in the stiff regime. To do so, we fix the
time step $h=1$, and choose $t_{\rm stop}$ to be less than $h$. The
other parameters are the same as in the previous test
($\epsilon=1, w_0=1$). We evolve the system to $t_e=10$, using
$t_{\rm stop}=0.2$ and $t_{\rm stop}=0.02$. In Figure
\ref{fig:collision2} we show the time evolution of particle velocity
from the semi-implicit and the fully-implicit integrators. We see
that with the fully-implicit integrator, the particle velocity
(thus the gas velocity, due to momentum conservation) rapidly drops
to zero, and smaller $t_{\rm stop}$ leads to faster damping [actually
$v^{(n+1)}\sim2(t_{\rm stop}/h)^2v^{(n)}$]. Because $t_{\rm stop}$ is
not resolved, the damping rate is slower than theoretical values,
but still is rapid as relative velocity $v$ drops several orders of
magnitude in each time step. With the semi-implicit integrator,
however, the particle velocity undergoes damped oscillation.
The smaller $t_{\rm stop}$ is, the slower the particle velocity damps
[actually $v^{(n+1)}\sim-(1-4t_{\rm stop}/h)v^{(n)}$], opposite to the
trend found in the fully-implicit integrator. In the limit
$t_{\rm stop}=0$, the particle velocity never dies away. The behavior
of the semi-implicit integrator in this regime exactly resembles a
damped harmonic oscillator, which makes it difficult for particles to
get rid of extra relative velocity with respect to gas. This situation
is problematic because if the gas itself undergoes some other
oscillation with a similar frequency, the amplitude of the oscillator
may be quickly amplified rather than damped due to (numerical) resonant
interactions. Therefore, for stability considerations, we use the
fully-implicit integrator when $t_{\rm stop}\lesssim h$.

\subsection[]{Streaming Instability in the Linear Regime}\label{ssec:linSI}

The most stringent test of our code is the streaming instability
linear growth rate test using eigenmodes provided by YJ07. These
eigenmodes are built from the Nakagawa-Sekiya-Hayashi (NSH)
equilibrium [\citealp{NSH86}; also see equation (7) of YJ07].
The NSH equilibrium refers to the equilibrium between solids and
gas in unstratified Keplerian disks, in which gas is partially
supported by a radial pressure gradient. Establishing the NSH
equilibrium numerically requires exact balance among all force terms.
With the help of the particle advection scheme as well as careful
treatment of predictor and corrector step momentum feedback (see
\S\ref{sec:hybrid} and \S\ref{sec:integrator} for details), we
are able to establish the {\it exact} NSH equilibrium in our code
(i.e., net force on both the gas and particles is zero to machine
precision). The eigenmodes of the streaming instability are
characterized by dimensionless wave numbers
$K_x\equiv k_x\eta v_k/\Omega$ and $K_z\equiv k_z\eta v_k/\Omega$.
Taking $\eta v_k=0.05c_s$ and setting the box size to be one
wavelength, we have $K_x=0.05(c_s/\Omega)(2\pi/L_x)$, and
similarly in the $z$ direction. We fix the box size to be $L_x=L_z=2$,
therefore, $K_x$ (and $K_z$) sets the sound speed.
We construct eigenvectors of density and velocity perturbations
using formula (10) and Table 1 in YJ07. We use one particle
per cell and each particle is located at cell center initially. In
order to generate particle density profile of
$\rho_d=1+A\cos{k_xx}\cos{k_zz}$ (we take $A=10^{-6}$), we shift
particle positions in the radial ($x$) direction, with the amount of
shift proportional to $\cos{k_zz}$.

\begin{figure*}
    \centering
    \subfigure[Test ``linA"]{
    \includegraphics[width=150mm,height=65mm]{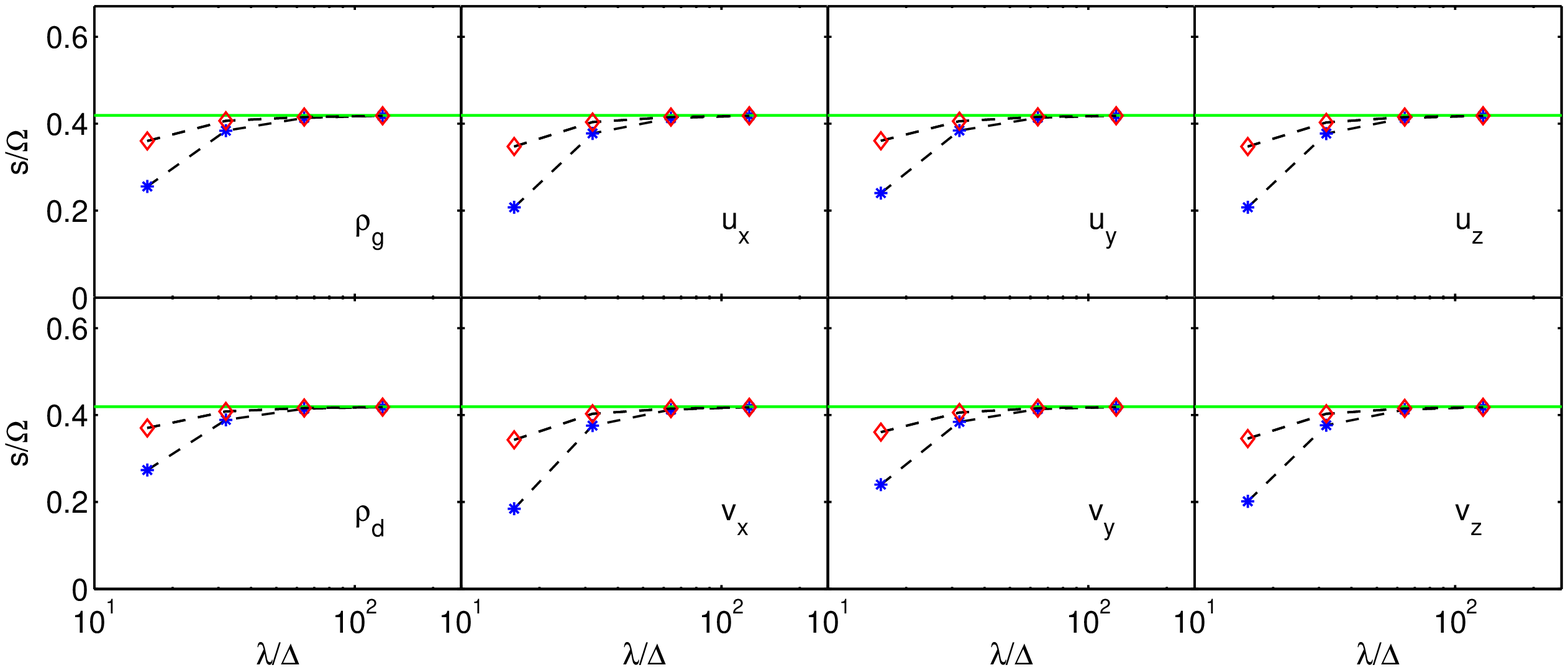}}
    \subfigure[Test ``linB"]{
    \includegraphics[width=150mm,height=65mm]{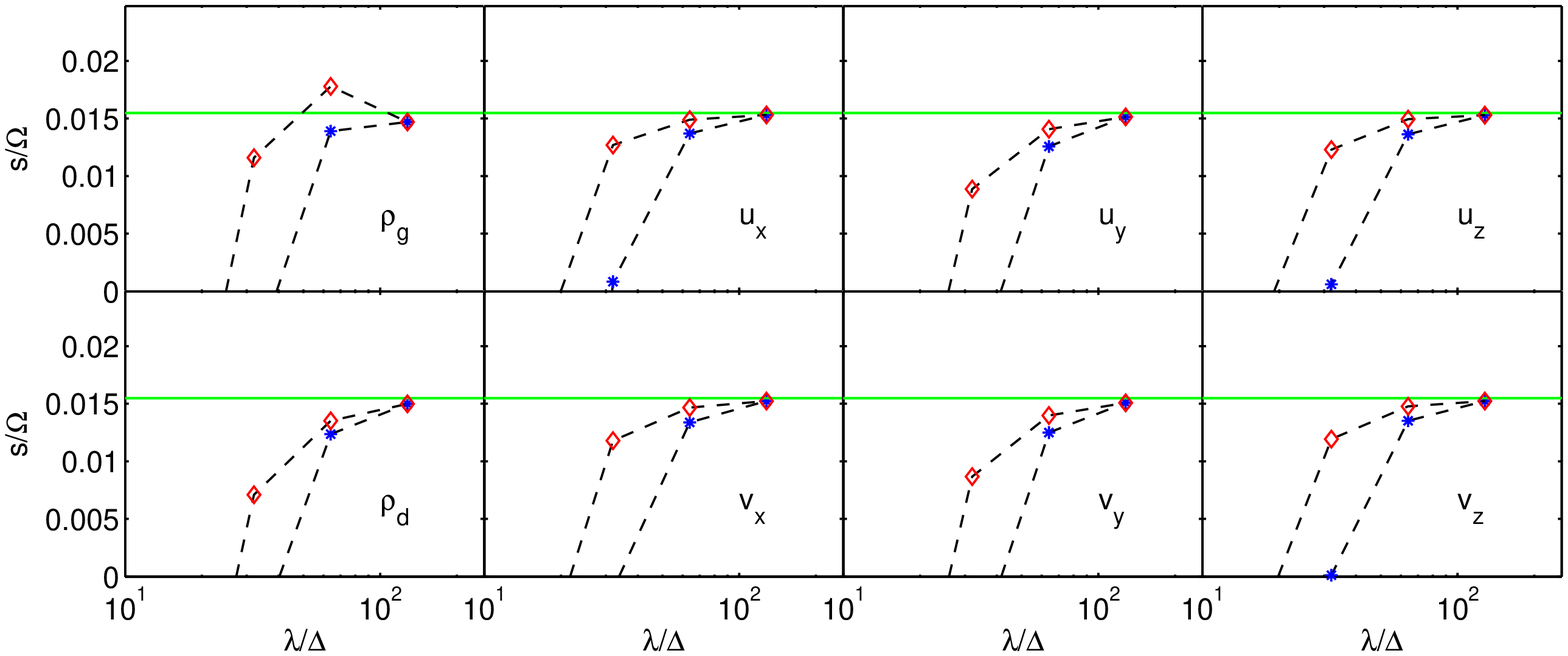}}
  \caption{Measured growth rate of the streaming instability eigenmodes
  ``linA" (a) and ``linB" (b) adopted from \citet{YoudinJohansen07} using
  the semi-implicit integrator. Solid line marks the theoretical growth rate,
  and dashed lines with symbols label the measured growth rate as a
  function of grid resolution (number of cells per wavelength). Asterisks
  correspond to evolution using fixed CFL number 0.8, diamonds
  correspond to fixed time step (set by CFL=0.8 with 128 cells).}\label{fig:streaminglin}
\end{figure*}

The test suite in YJ07 aims at measuring the numerical growth
rate of the streaming instability eigenmodes as a function of
grid resolution. It consists of two problems, both using
$\tau_s=0.1$ particles.  The test ``linA" has particle to gas mass
ratio $\epsilon=3$ and $K_x=K_z=30$. The predicted growth rate is
$s=0.4190204\Omega$. The second test ``linB" has $\epsilon=0.2$ and
$K_x=K_z=6$. This test is more challenging because the mode grows
very slowly, with $s=0.015476\Omega$.\footnote{Also, due to a smaller
value of $K$, the sound speed is smaller than the ``linA" test, thus larger
time step.} We consider grid resolutions of 16, 32, 64 and 128 cells per
wave length. Figure \ref{fig:streaminglin} shows our test results using the
semi-implicit integrator. In each test, we conduct two set of runs, one with
fixed Courant-Friedrichs-Lewy number CFL=0.8, and the other with a
fixed time step (or CFL=0.1, 0.2, 0.4, 0.8 for each resolution). For
reference, the time step with lowest resolution (16 cells per wavelength)
and CFL=0.8 is $\Omega h=2.6\times10^{-4}$ for ``linA" and
$\Omega h=1.3\times10^{-3}$ for ``linB".  Both semi-implicit and
fully-implicit methods are considered, and the results are almost
identical (since the time step is much smaller than $t_{\rm stop}$). We
see that our code converges very well to the predicted growth rate
in the run ``linA". Small time step helps with convergence. For the
more challenging `linB" test, we get slower convergence. In particular,
the mode does not grow for 32-cell resolution using CFL=0.8. Better
temporal resolution again improves the performance.

These test results show that with CFL=0.8, about 64 cells are needed to see
growth (e.g., test linB), although 16 cells seem to be sufficient to capture the
most unstable modes (e.g., test linA). Viewed from the results, our measured
linear growth rates are similar to (and even closer to semi-analytic values
than) those of \citet{Balsara_etal09} and \citet{Miniati10}, who employ similar
MHD code but different hybrid schemes. Our method is less accurate than
that presented in YJ07, which is not surprising given the fact that our
code is second order accurate (both spatial and temporal) while the Pencil
code used by YJ07 is sixth order spatial and third order temporal accurate
for smooth flows, such as in this test.

The test problems ``linA" and ``linB" provided in YJ07 both adopt relatively
large particle stopping time $\tau_s=0.1$. Since we will perform SI
simulations on a size distribution of particles with stopping time down to
$10^{-3}$, additional code test is needed to justify the ability of our code.
In Appendix \ref{app:lin}, we provide two more linear growth tests for
particles with $\tau_s=10^{-2}$ and $\tau_s=10^{-3}$. Numerical
convergence to the analytical growth rate is again achieved, although
slightly higher resolution is required to reach the convergence.

\section[]{Convergence of Streaming Instability}\label{sec:converge}

As an application of our hybrid code, we study the streaming
instability in the non-linear regime. Such calculations have been
performed and comprehensively analyzed by JY07. In this paper,
we focus on the numerical convergence of physical properties in
the saturated state, which was not fully explored in JY07.
Understanding the convergence properties of hydrodynamic simulations
of the streaming instabilities is important before adding more
complicated physics (such as MHD), and interpreting the reliability
of such simulations.

We study the numerical convergence using 2D simulations in the
$x-z$ (radial-vertical) plane in which both the grid resolution and
the number of (super-) particles vary. By using 2D simulations, we
can explore a wide range of resolutions that is not possible in fully
3D simulations. Moreover, the results of 2D and 3D simulations are
similar (JY07), thus one expects the numerical convergence properties
of 2D simulations are similar to those in 3D as well.

The two most significant effects of the streaming instability are
the concentration of particles into dense clumps, and the generation
of turbulence and/or waves. The former can be characterized by the
probability distribution function (PDF) of particle densities, while
the latter can be characterized by the turbulent velocity, diffusion
coefficient and momentum flux. We investigate the numerical
convergence of these two aspects in \S\ref{ssec:concentration}
and \S\ref{ssec:turbulence} respectively.

\subsection[]{Simulation Setup}\label{ssec:setup}

We perform simulations of the streaming instability using an isothermal
equation of state and neglecting vertical gravity. In this case, the
properties of the streaming instability is largely determined by two
parameters, the particle stopping time $\tau_s$ and the overall mass
ratio between particles and gas $\epsilon$. The drift velocity
$\eta v_K$ does not serve as an independent variable, but it sets the
length scale of the problem, that is, all length scales can be measured
in units of $\eta r\equiv\eta v_K/\Omega$. The isothermal sound speed
$c_s$ is also a model parameter, but it is much less important because
the gas motion is almost incompressible (\citealp{YoudinGoodman05},
JY07). We adopt $\eta v_K/c_s=0.05$ throughout this paper, which roughly
matches the value expected at 1AU in the MMSN model.

As shown by JY07, there are two basic ``modes" to the non-linear
saturation of the streaming instability. For marginally coupled and
larger particles ($\tau_s\gtrsim1$), the instability develops for a wide
range of $\epsilon$, around 1. In the saturated state, particles
concentrate into long stripes, which are mostly aligned with the vertical
direction and are slightly tilted in the radial direction. For tightly
coupled particles ($\tau_s\ll1$), prominent instability develops only
when $\epsilon\gtrsim1$, and the turbulent state consists of large voids
with narrow particle streams. The turbulence is much weaker than the
previous case (see Figure 2 and Figure 5 of JY07 for each of the two
``modes", also see our Figure \ref{fig:streaming} below).
We have also performed a series of simulations with a wide range of
parameters ($\tau_s$ and $\epsilon$), and have reproduced all the results
in JY07. We use the semi-implicit integrator in all the tests in this section,
and results from the other two integrators are similar, since the time step is
at least 10 times smaller than $t_{\rm stop}$ in all test runs.

To study convergence, we choose two test problems that are
representative of the two saturation ``modes" described by JY07
\footnote{We have also studied the non-linear behavior of the SI
for more strongly coupled particles with different resolutions. Briefly
speaking, finer resolution is needed to capture the SI as $\tau_s$
gets smaller. Particle clumping becomes weaker as $\tau_s$
decreases and reduces to modest overdensities as
$\tau_s\leq0.01$ (with $\epsilon\geq1$). The results are in qualitative
agreement with the linear analysis by \citet{YoudinGoodman05}. In
this paper, however, we focus on the non-linear SI tests in JY07
for conciseness, the main conclusions are also applicable to the case
with more strongly coupled particles.}.
The first problem is the same as their run AB. The run parameters
include: $\tau_s=0.1$, $\epsilon=1.0$, $L_x=L_z=2\eta r$. The
second problem uses parameters from their run BA, with $\tau_s=1.0$,
$\epsilon=0.2$, $L_x=L_z=40\eta r$. We choose the ``standard" grid
resolution to be $256^2$, and $N_{\rm pc}=9$ particles per grid cell.
The typical time step is given by $h\sim{\rm CFL}(L_x/N)/c_s$ where
$N$ is the number of grid cells in one direction. For the standard grid
resolution and ${\rm CFL}=0.8$, we have $\Omega h\approx3\times10^{-4}$
for run AB, and $\Omega h\approx6\times10^{-3}$ for run BA.
We then conduct a series of runs for each problem. First, we change
the grid resolution from $64^2$ to $2048^2$, with the number of
particles per cell fixed at 9. Note that increasing spatial resolution
at fixed ${\rm CFL}$ number is accompanied by the increase of
temporal resolution, which is also important for driving to convergence,
as we discussed in \S\ref{ssec:linSI}. Secondly, we fix the grid resolution
to be $256^2$, and change the number of particles per cell,
$N_{\rm pc}=1, 4, 9, 16$ and $25$. In each of
these runs, particles are initialized with random positions within
the simulation box, with velocities taken from the NSH equilibrium.
All these runs are tabulated in the first three columns of Table
\ref{tab:turbulence}, where each variation of the AB or BA runs is
assigned with a run number. Our fiducial runs are AB3 and BA3, while
our runs AB9 and BA9 have exactly the same numerical parameters as
that in JY07.

The AB runs are fully saturated after about $16\Omega^{-1}$ and the
BA runs do not saturate until about $160\Omega^{-1}$. We choose the
saturation time for the two runs to be some later time
$T_s^{(AB)}=30\Omega^{-1}$, $T_s^{(BA)}=300\Omega^{-1}$, after which
we start collecting data from the simulations and perform measurements.
In Figure \ref{fig:streaming}, we show images of the particle density
with different grid resolution ($64^2$, $256^2$, $1024^2$ from top to
bottom) for runs AB (left) and BA (right) at saturation ($t=T_s$). Our
standard runs (middle panels) resemble the last panels of Figure 4 and
Figure 5 in JY07. For the BA run series, we see that the bulk patterns
of particle density do not vary much with grid resolution, although more
and more detailed features are revealed in the higher resolution
simulations. For the AB run series, all three grid resolutions show
cavitation and particle streams as described in JY07, however, the scale
of particle clumping becomes smaller as one uses finer resolutions. A
resolution of $256^2$ appears to be necessary to capture the prominent
features in the particle density pattern. The results of these runs will
serve as the starting point of our more quantitative study of numerical
convergence in the next two subsections, using physical quantities
averaged over the period between $T_s$ and $T_e$, where
$T_e^{(AB)}=80\Omega^{-1}$ for AB runs and $T_e^{(BA)}=1500\Omega^{-1}$
for BA runs.

\begin{figure*}
  \vspace*{-4mm}
  \begin{center}
      \includegraphics[width=7.0cm,height=6.5cm]{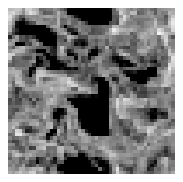}
      \includegraphics[width=7.0cm,height=6.5cm]{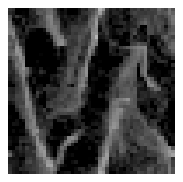}
      \includegraphics[width=7.0cm,height=6.5cm]{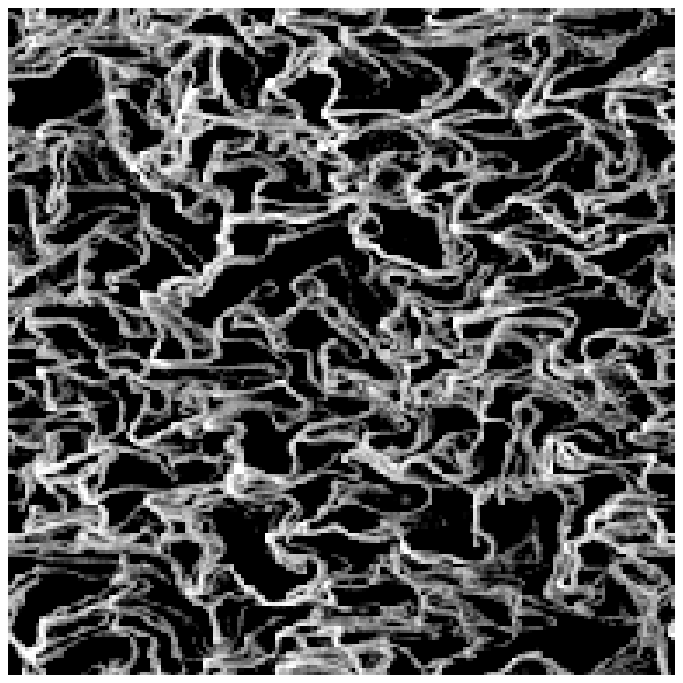}
      \includegraphics[width=7.0cm,height=6.5cm]{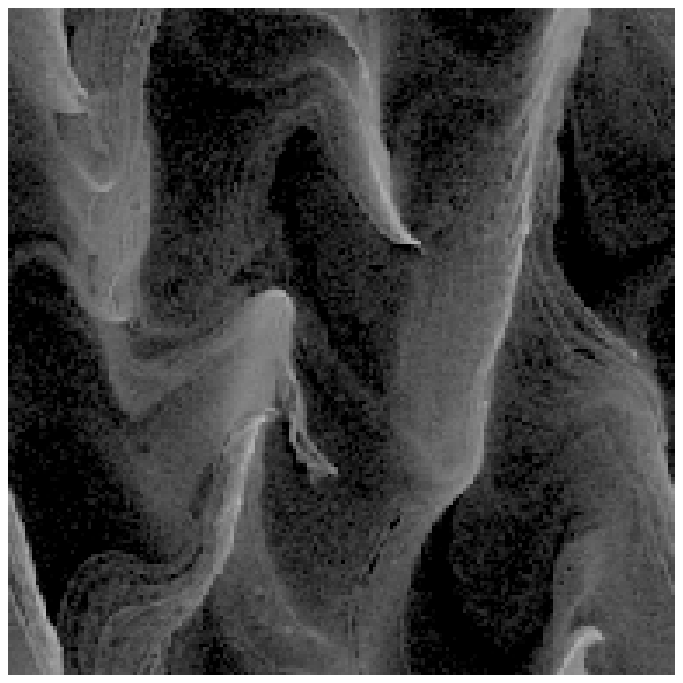}
      \includegraphics[width=7.0cm,height=6.5cm]{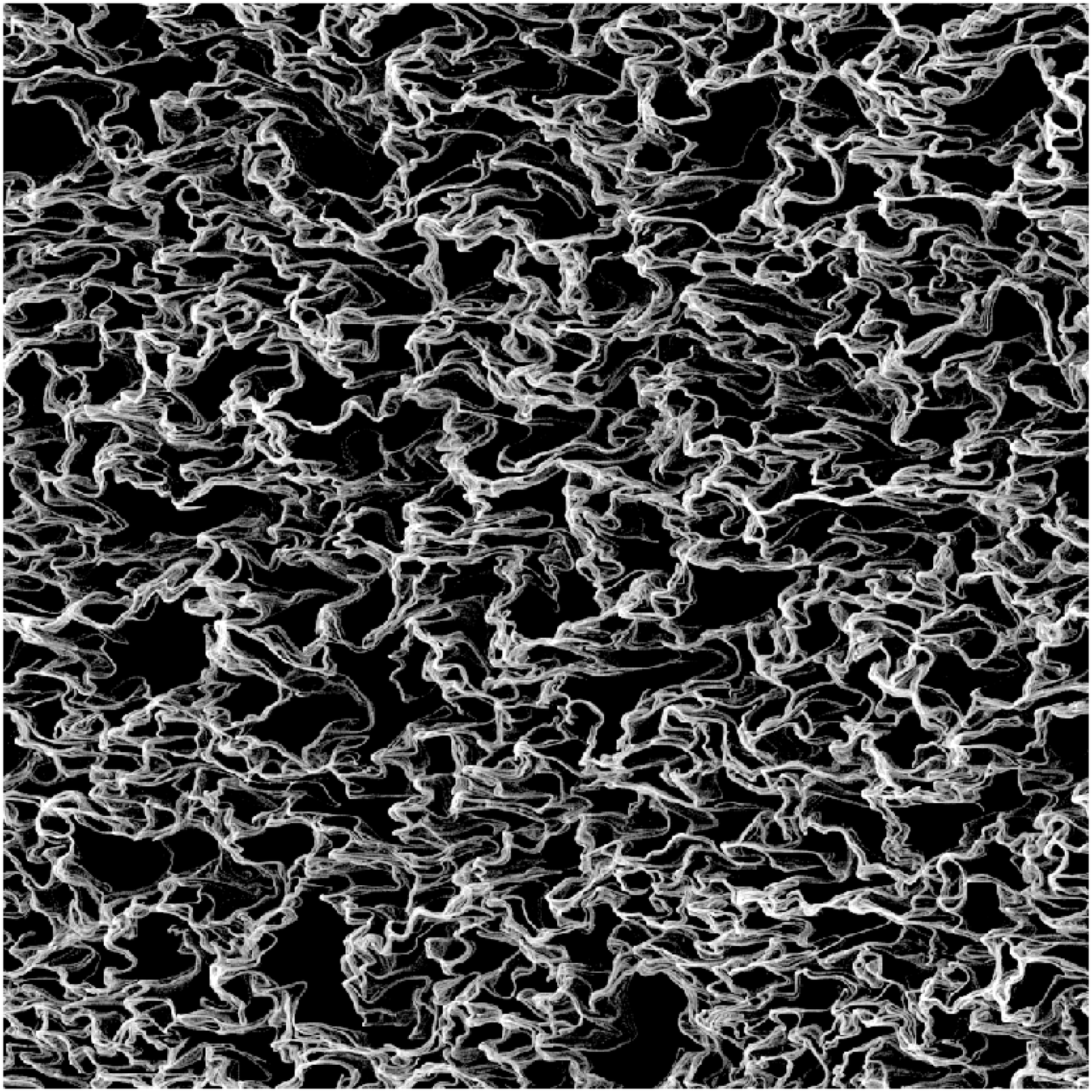}
      \includegraphics[width=7.0cm,height=6.5cm]{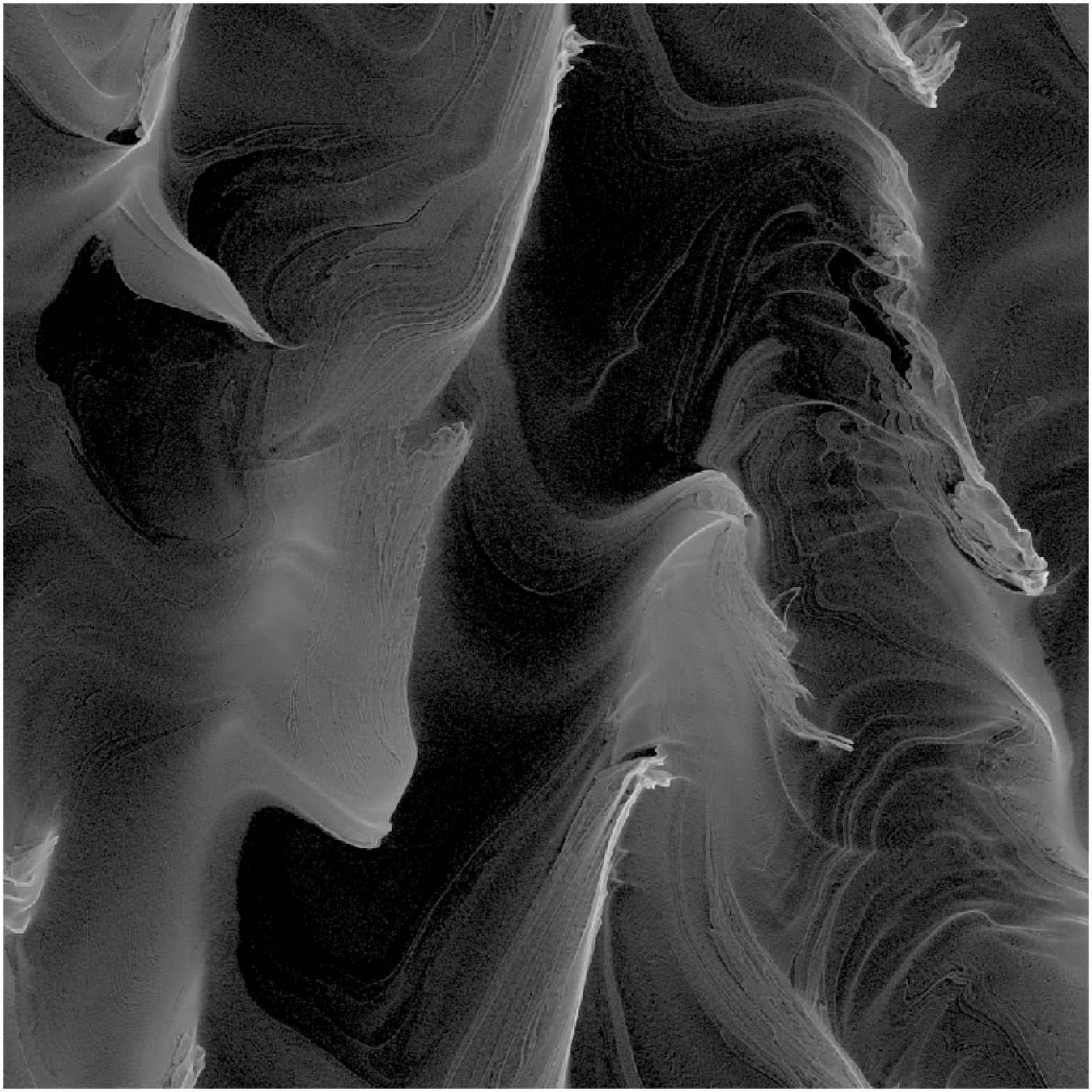}
  \end{center}
  \caption{Snapshots of particle densities in the $x-z$ plane of run AB
  (left) and BA (right) at saturation of the streaming instability $T=T_s$
  with different grid resolutions: $64^2$ (top), $256^2$ (middle) and
  $1024^2$ (bottom). All simulations use $9$ particles per cell. Particle
  densities are shown with log scale, ranging between $0.1$ to $10$ of the
  gas density for the AB runs, and between $0.01$ to $100$ for the BA runs.
  White regions indicate high density. The size of the box is
  $L_x=L_z=2\eta r$ for run AB, and $L_x=L_z=40\eta r$ for run BA.}\label{fig:streaming}
\end{figure*}

\subsection[]{Particle Concentration}\label{ssec:concentration}

\begin{figure*}
  \vspace*{-4mm}
  \begin{center}
    \includegraphics[width=8.0cm,height=6.5cm]{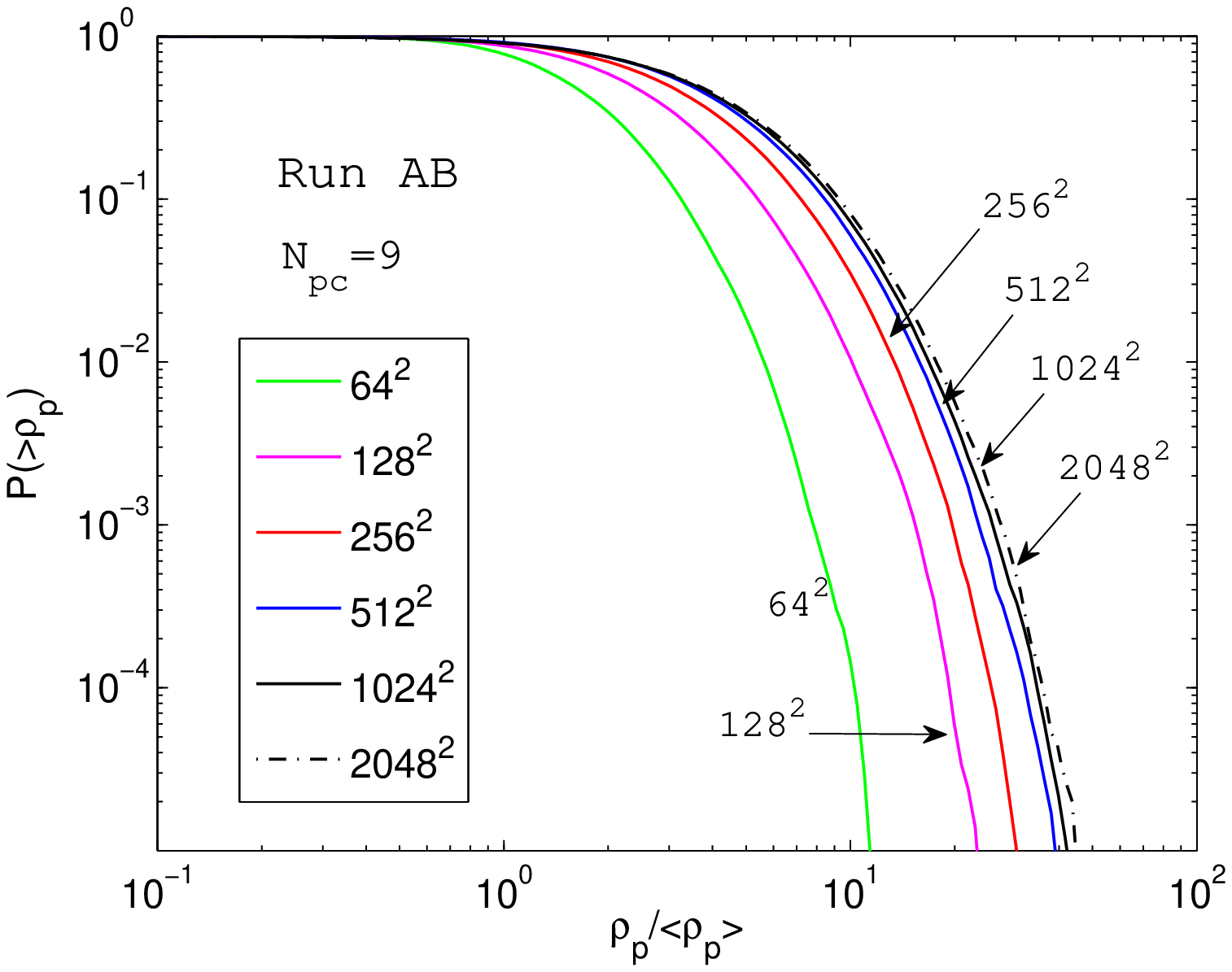}
    \includegraphics[width=8.0cm,height=6.5cm]{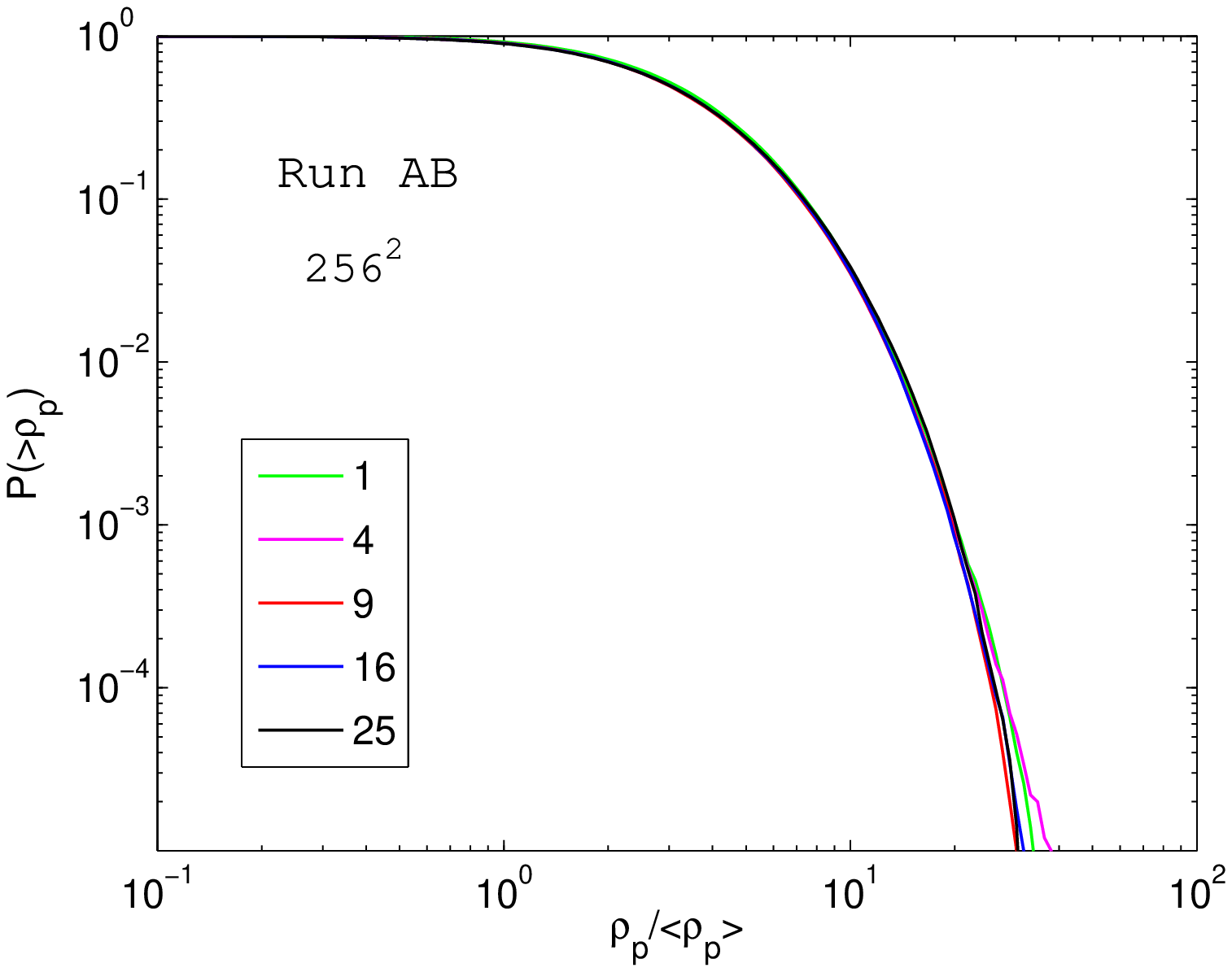}
    \includegraphics[width=8.0cm,height=6.5cm]{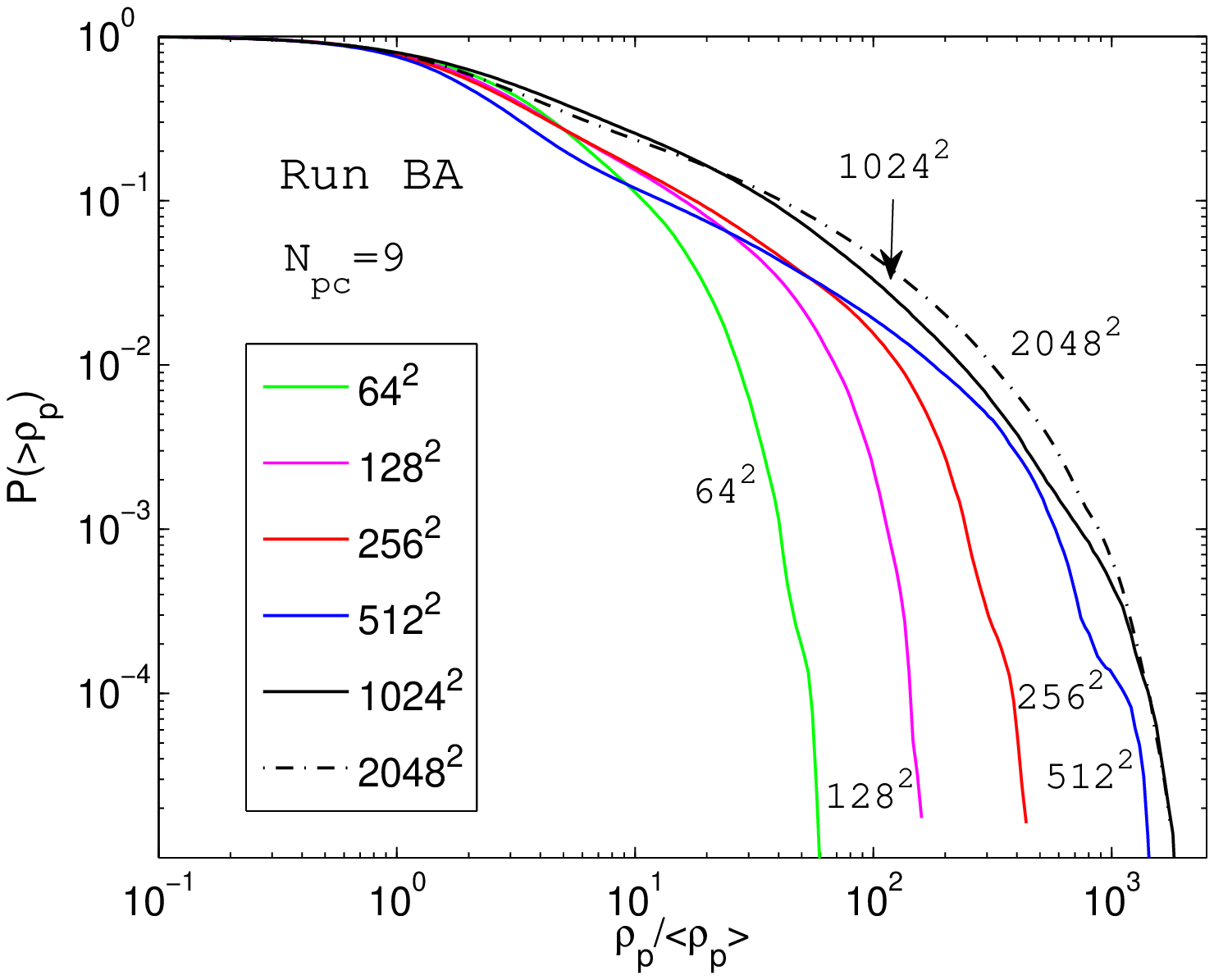}
    \includegraphics[width=8.0cm,height=6.5cm]{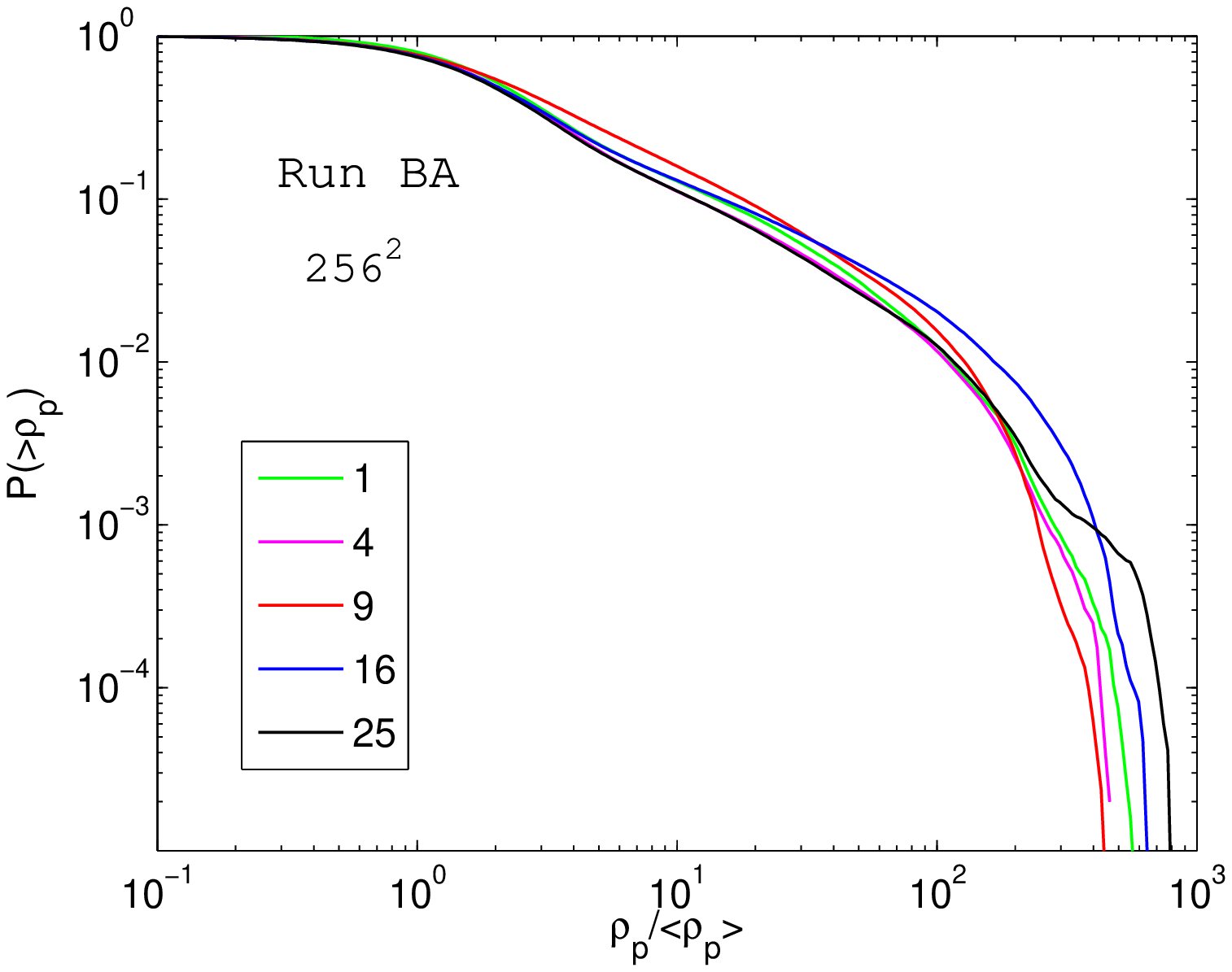}
  \end{center}
  \caption{The cumulative probability distribution function of particle
  density. Top and bottom panels show results from the AB and BA runs
  respectively. On the left panels, number of particles per cell
  $N_{\rm pc}$ is fixed to 9. Green, magenta, red, blue and black
  curves label results from $64^2$, $128^2$, $256^2$, $512^2$ and $1024^2$
  grid resolutions respectively. The dash-dotted curve shows the results from
  $2048^2$ resolution run for comparison.
  On the right panels, grid resolution is fixed to be $256^2$, while
  $N_{\rm pc}$ varies to be $1$, $4$, $9$, $16$ and $25$, as labeled by
  green, magenta, red, blue and black curves.}\label{fig:pdf}
\end{figure*}

Probably the most important property of the streaming instability
is its ability to concentrate particles. In JY07, it was shown that
the maximum particle density resulting from the streaming instability
can reach as high as $10^3$ times background particle density. Such
high densities are sufficient to make the clumps gravitationally
bound, promoting the formation of planetesimals in PPDs
\citep{Johansen_etal07}. The concentration of particles is best
demonstrated by the cumulative probability distribution function (PDF)
of particle densities. It measures the fraction of particles whose
ambient particle density exceeds a given value. In Figure \ref{fig:pdf}
we show the particle density distribution from the series of AB and BA
runs. Note that in the horizontal axis, we have normalized particle
density to the averaged particle density in the simulation box.

The ability for our code to handle the particle clump is reflected in the
stiffness parameter $\chi$ defined in equation (\ref{eq:stiffness}). We
have tested that the maximum value of $\chi$ in our AB runs is about
$\chi_{\rm max}\sim0.2$ in $64^2$ resolution, and it monotonically
drops to $\chi_{\rm max}\sim0.004$ in $2048^2$ resolution. For the
BA runs, $\chi_{\rm max}$ stays less than $0.5$ for most of the time
and for all resolutions, but reaches as large as $2.0$ in a few transients
in the highest resolution run. The duration of the transients is so short
that their contribution to the PDF is far below $10^{-5}$ and is not visible
in our Figure \ref{fig:pdf}. These facts show that the particle clumps are
properly handled in our code and the obtained PDFs are not affected
by the possible stiffness in the dense clumps.

Our results agree quantitatively with Figure 11 of JY07 (our AB9 and
BA9 $256^2$ runs resolution with $N_{\rm pc}=25$). The PDFs calculated
from our AB runs are very robust, in the sense that using longer period
of averaging or using different initial random seeds do not produce any
visible changes in the plot. For the BA runs, there can be some
uncertainties in the calculation of the PDFs, mostly because that there
are only a few dense clumps in the simulation box, the long term
stochastic evolution of these big clumps makes the calculated PDFs more
or less dependent on the period of time averaging. Nevertheless, we have
averaged the PDFs over as long as nearly 200 orbits, which is about 170
times the correlation time of the clumps (which is about $7\Omega^{-1}$,
see Figure 13 of JY07), such uncertainty is expected to be small and does
not affect the main features to be discussed below.

The left panels of Figure \ref{fig:pdf} show the effect of grid resolution
on the density PDFs: higher grid resolution leads to stronger particle
clumping. Both the maximum particle density is higher, and the number of
particles residing in dense clumps is larger at higher resolutions. For the AB
runs, the PDFs from the $512^2$ resolution run is almost identical to higher
resolution runs, indicating convergence. For the BA runs, the small number
of dense clumps in the simulation box makes convergence more difficult
by averaging over a finite time period. Nevertheless, viewing from the PDF
curves, it appears that convergence is finally reached at $1024^2$ resolution.

We have also studied the numerical convergence with number of
particles in the simulation box. On the right panels of Figure
\ref{fig:pdf}, we see that for both runs AB and BA, the PDFs depend very
weakly on $N_{\rm pc}$. In the AB runs, the five curves almost overlap with
each other. In BA runs, however, there is larger scatter. Again, such scatter
is most likely due to the small number of clumps. In fact, we have compared
the BA run PDFs between short ($T_e-T_s=500\Omega^{-1}$) and long
($T_e-T_s=1200\Omega^{-1}$, shown in the figure) averaging period. The
PDFs from longer averaging converges show better convergence\footnote{
In particular, the PDF from our run BA3 ($N_{\rm pc}=9$) appears to have
substantial lower peak densities than the runs with other $N_{\rm pc}$s
when we average over shorter time period.}. We expect the scatter to
become smaller if we run BA series for much longer time. In all, we
conclude that one can use as small as only one particle per cell to
accurately capture the density distributions of the solids.

\subsection[]{Turbulence Properties}\label{ssec:turbulence}

The streaming instability generates turbulence which differs
significantly from the laminar state described by the NSH equilibrium.
The turbulence is accompanied by particle concentration, as discussed in
the previous subsection. Understanding the turbulent state is important
since it may strongly affect the settling process as well as the radial
transport of solids in the PPDs (JY07, \citealp{BaiStone10b}). Similar
to JY07, we have computed a number of physical quantities to characterize
the turbulence properties generated from the streaming instability,
including: 1) The turbulent Mach number ($Ma$) in three directions,
calculated from the root mean square of gas velocity fluctuations; 2)
Mean radial drift velocity of the particles, normalized by the NSH radial
drift velocity $\overline{v_x}/\overline{v_x}^{NSH}$; 3) Turbulent
diffusion coefficient for particles $D$, calculated using the method
outlined in \S5.3 of JY07; and 4) The time and spatial averaged Reynolds
stress ${\mathcal F}\equiv\rho_gu_x(u_y-v_K)$, divided by the Reynolds
stress in the NSH equilibrium (note the quantity we measure is different
from ${\mathcal F}_{\mathcal{L},x}$ in JY07). Properties (1), (2) and (4)
are calculated directly from spatial and time averaging from the entire
simulation box. The $1\sigma$ uncertainties are estimated from the
standard deviation in the time sequence. In the calculation of the
diffusion coefficient, we measure the standard deviation of the distance
traveled by tracer particles at different time intervals $\sigma_x(\Delta t),
\sigma_z(\Delta t)$. We then fit the diffusion coefficient using
$\sigma^2_{x,z}(\Delta t)=2D_{x,z}\Delta t$. The maximum $\Delta t$ has
been chosen to be $32\Omega^{-1}$ for AB runs, $320\Omega^{-1}$ for BA runs.
The uncertainty of the diffusion coefficient is directly given by the linear
regression estimate. The results are summarized in Table \ref{tab:turbulence}.

From Table \ref{tab:turbulence} we see that the gas flow properties, namely
the turbulent Mach number and the Reynolds stress, depend relatively weakly
on the grid resolution and number of particles in the simulation box. This
trend is also true for the radial drift velocity of the particles. In the AB
series of runs, all these quantities agree with each other once the grid
resolution reaches $256^2$. Smaller resolution runs produce somewhat different
values, but with larger fluctuations. Note that in Figure \ref{fig:streaming},
the typical size of particle stripes from the $1024^2$ run (AB5) is apparently
smaller than that from the $256^2$ run (AB3). However, the statistical
properties of gas flow from these two runs are indistinguishable. In this
sense, $256^2$ resolution is sufficient to capture the essential turbulent
properties for the AB runs. For BA run series, the measurements from even the
lowest resolution run $64^2$ agree very well with other higher resolution runs.
Such numerical convergence behavior is very different from the convergence of
particle concentration properties discussed in \S\ref{ssec:concentration}.

The measurements of the diffusion coefficient in the radial and vertical
directions, however, show larger variations among different runs for both
AB and BA series. Such variations show up not only between different grid
resolutions, but also at fixed resolution, varying the number of particles in
the simulation box also makes a difference. The differences even exceed the
$1\sigma$ uncertainties in a number of cases. Moreover, there is no
systematic trend on such variances, especially in the BA run series.
In the AB runs, it appears that towards higher resolution, the diffusion
coefficient becomes smaller. Given the fact that bulk flow properties converge
well at modest resolutions, the variation of diffusion coefficient
measured from different runs, especially those with grid resolution no
less than $256^2$, may be taken as uncertainties from the numerical simulations.
The uncertainties in the measured diffusion coefficient would be about 20\%,
for both AB and BA runs.

\begin{landscape}
\begin{table*}
\caption{Turbulence Properties.}\label{tab:turbulence}
\begin{center}
\begin{tabular}{ccccccccccc}\hline\hline
 Run & Resolution & $N_{\rm pc}$ & $Ma_x\ ^1$ & $Ma_y\ ^1$ & $Ma_z\ ^1$ &
 $\overline{v_x}/\overline{v_x}^{NSH}\ ^2$ & $D_x\ ^3$ & $D_z\ ^3$
 & ${\mathcal F}_{\rm Re}/{\mathcal F}^{\rm NSH}_{\rm Re}\ ^4$\\\hline
 AB1 & $64^2$   & 9 & $0.97(15)\times10^{-2}$ & $2.44(15)\times10^{-2}$ & $0.81(15)\times10^{-2}$
                    & 1.40(80) & $2.64(08)\times10^{-5}$ & $7.01(29)\times10^{-5}$ & 1.72(79)\\
 AB2 & $128^2$  & 9 & $1.23(07)\times10^{-2}$ & $2.45(07)\times10^{-2}$ & $0.98(07)\times10^{-2}$
                    & 1.88(27) & $4.61(05)\times10^{-5}$ & $9.50(63)\times10^{-5}$ & 2.38(30)\\
 AB3 & $256^2$  & 9 & $1.24(04)\times10^{-2}$ & $2.47(04)\times10^{-2}$ & $0.78(04)\times10^{-2}$
                    & 2.14(07) & $5.12(04)\times10^{-5}$ & $3.01(06)\times10^{-5}$ & 2.67(12)\\
 AB4 & $512^2$  & 9 & $1.25(03)\times10^{-2}$ & $2.47(03)\times10^{-2}$ & $0.78(03)\times10^{-2}$
                    & 2.21(06) & $4.58(04)\times10^{-5}$ & $2.89(07)\times10^{-5}$ & 2.69(10)\\
 AB5 & $1024^2$ & 9 & $1.24(03)\times10^{-2}$ & $2.47(03)\times10^{-2}$ & $0.76(03)\times10^{-2}$
                    & 2.22(05) & $4.37(06)\times10^{-5}$ & $2.28(02)\times10^{-5}$ & 2.68(09)\\
 AB6 & $256^2$  & 1 & $1.23(05)\times10^{-2}$ & $2.46(05)\times10^{-2}$ & $0.76(05)\times10^{-2}$
                    & 2.13(08) & $5.11(06)\times10^{-5}$ & $2.76(07)\times10^{-5}$ & 2.67(16)\\
 AB7 & $256^2$  & 4 & $1.24(05)\times10^{-2}$ & $2.46(05)\times10^{-2}$ & $0.80(05)\times10^{-2}$
                    & 2.13(08) & $5.16(03)\times10^{-5}$ & $3.15(10)\times10^{-5}$ & 2.67(15)\\
 AB3 & $256^2$  & 9 & $1.24(04)\times10^{-2}$ & $2.47(04)\times10^{-2}$ & $0.78(04)\times10^{-2}$
                    & 2.14(07) & $5.12(04)\times10^{-5}$ & $3.01(06)\times10^{-5}$ & 2.67(12)\\
 AB8 & $256^2$ & 16 & $1.26(06)\times10^{-2}$ & $2.47(06)\times10^{-2}$ & $0.81(06)\times10^{-2}$
                    & 2.16(09) & $5.14(04)\times10^{-5}$ & $3.29(07)\times10^{-5}$ & 2.71(16)\\
 AB9 & $256^2$ & 25 & $1.26(05)\times10^{-2}$ & $2.46(05)\times10^{-2}$ & $0.83(05)\times10^{-2}$
                    & 2.15(08) & $5.38(05)\times10^{-5}$ & $3.25(08)\times10^{-5}$ & 2.72(16)\\\hline
 BA1 & $64^2$   & 9 & $1.19(08)\times10^{-2}$ & $1.84(08)\times10^{-2}$ & $3.88(08)\times10^{-2}$
                    & 0.65(07) & $2.18(06)\times10^{-3}$ & $1.01(04)\times10^{-2}$ & 0.59(07)\\
 BA2 & $128^2$  & 9 & $1.20(10)\times10^{-2}$ & $1.87(10)\times10^{-2}$ & $4.03(10)\times10^{-2}$
                    & 0.65(06) & $2.36(06)\times10^{-3}$ & $1.37(06)\times10^{-2}$ & 0.61(06)\\
 BA3 & $256^2$  & 9 & $1.21(11)\times10^{-2}$ & $1.81(11)\times10^{-2}$ & $3.88(11)\times10^{-2}$
                    & 0.66(07) & $2.30(09)\times10^{-3}$ & $0.97(05)\times10^{-2}$ & 0.62(08)\\
 BA4 & $512^2$  & 9 & $1.16(11)\times10^{-2}$ & $1.83(11)\times10^{-2}$ & $3.93(11)\times10^{-2}$
                    & 0.70(06) & $2.29(11)\times10^{-3}$ & $1.39(05)\times10^{-2}$ & 0.67(06)\\
 BA5 & $1024^2$ & 9 & $1.36(09)\times10^{-2}$ & $1.88(09)\times10^{-2}$ & $3.89(09)\times10^{-2}$
                    & 0.60(06) & $2.20(05)\times10^{-3}$ & $1.16(08)\times10^{-2}$ & 0.56(06)\\
 BA6 & $256^2$  & 1 & $1.17(12)\times10^{-2}$ & $1.86(12)\times10^{-2}$ & $4.00(12)\times10^{-2}$
                    & 0.67(08) & $2.71(11)\times10^{-3}$ & $1.44(05)\times10^{-2}$ & 0.64(08)\\
 BA7 & $256^2$  & 4 & $1.12(05)\times10^{-2}$ & $1.80(05)\times10^{-2}$ & $3.92(05)\times10^{-2}$
                    & 0.71(04) & $1.87(03)\times10^{-3}$ & $1.23(05)\times10^{-2}$ & 0.68(04)\\
 BA3 & $256^2$  & 9 & $1.21(11)\times10^{-2}$ & $1.81(11)\times10^{-2}$ & $3.88(11)\times10^{-2}$
                    & 0.66(07) & $2.30(09)\times10^{-3}$ & $0.97(05)\times10^{-2}$ & 0.62(08)\\
 BA8 & $256^2$ & 16 & $1.16(10)\times10^{-2}$ & $1.83(10)\times10^{-2}$ & $4.00(10)\times10^{-2}$
                    & 0.69(05) & $2.15(06)\times10^{-3}$ & $1.35(06)\times10^{-2}$ & 0.66(05)\\
 BA9 & $256^2$ & 25 & $1.14(08)\times10^{-2}$ & $1.81(08)\times10^{-2}$ & $3.89(08)\times10^{-2}$
                    & 0.70(06) & $2.23(08)\times10^{-3}$ & $1.32(05)\times10^{-2}$ & 0.67(06)\\
\hline\hline
\end{tabular}
\end{center}
The number in parenthesis quotes the $1\sigma$ uncertainty of the last two digits.
See \S\ref{ssec:turbulence} for details.

$^1$ Mach number in the radial, azimuthal and vertical directions.

$^2$ Radial drift velocity of particles, normalized by the NSH value.

$^3$ Turbulent diffusion coefficient of the particles in the radial and vertical directions.

$^4$ Mean Reynolds stress of the gas, normalized by the NSH value.
\end{table*}
\end{landscape}

\section[]{Summary and Conclusion}\label{sec:conclusion}

We have presented the implementation of a hybrid particle-gas
scheme in the grid-based Athena MHD code. The particle and gas
are assumed to be coupled aerodynamically, and a size distribution
of particle species with different stopping times is allowed. Our
implementation is extendable to include gravitational coupling,
which is left for future work. The main purpose for the code
development is to study the dynamics of gas and solids in
protoplanetary disks (PPDs). The solid size range where the
aerodynamic coupling has significant effects in PPDs roughly spans
from millimeter to a few tens or hundreds meter sized bodies,
depending on the disk model and location, and this is the regime
most relevant for studying planetesimal formation \citep{ChiangYoudin10}.
In this paper we mainly address the numerical method and code
performance. In a forthcoming paper \citep{BaiStone10b}, we will
describe applications to PPDs.

The numerical algorithm of our hybrid code is based on a second-order
accurate predictor-corrector scheme. The algorithm is different from
other existing codes (YJ07,\citealp{Balsara_etal09,Miniati10}), and
is very simple and robust. Our implementation of particle-gas coupling
is fully conservative: backreaction from the particles to the gas is
treated carefully that the total momentum is conserved exactly. Our
hybrid code works well in non-stiff regime of particle-gas coupling,
as well as the stiff regime without significant particle mass loading
[i.e., the parameter $\chi$ defined in equation (\ref{eq:stiffness})
does not exceed order unity]. This is made possible by two implicit
particle integrators. These include a semi-implicit integrator,
generalized from the second order leap-frog integrator, which has much
better stability properties than any explicit methods, and which
preserves the geometric properties of particle orbit exactly in the
limit of zero drag force. In addition, a fully-implicit integrator is
designed for treating extremely stiff problem when particle stopping
time is much smaller than the simulation time step. We have extensively
tested our code performance, including the linear growth rate test of
the streaming instability. Subsequent test on the non-linear saturation
of the streaming instability further confirms that our code performs as
good as the higher order finite difference Pencil code, as the flow
properties measured from our simulations agree well with the results
by JY07.

We have also studied the numerical convergence of our method in the
non-linear regime of the streaming instability. We pick two
representative runs from JY07 and have performed a series of
simulations by varying grid resolutions and total number of
particles in the simulation box. We find that the convergence properties
in the non-linear regime is very different and more complicated than
those in the linear regime. The main conclusions drawn from this
convergence study are summarized below.

\begin{enumerate}
\item Requirement for numerical convergence strongly depends on
particle stopping time $\tau_s$. For $\tau_s=0.1$, about $128$ cells
per $\eta r$ is necessary for convergence. 

\item Equal number of particles and grid cells is sufficient for numerical
convergence.

\item Gas flow properties converge very well at modest grid resolution
and is not sensitive to number of particles used in the simulation.

\item Particle concentration properties converge at very high grid
resolution. Higher resolution leads to stronger clumping.

\item Particle diffusion properties depend on the numerical setup
in a subtle way, leading to about $\pm20\%$ uncertainties in the
measurements.
\end{enumerate}

These convergence tests provide useful information on the reliability
and uncertainty of this kind of hybrid simulations of the particle-gas
interaction. They also provide a guide on the choice of grid resolution
and number of particles for future simulations of similar and more
complicated problems. Although all the results are based on vertically
unstratified simulations, we may generalize these criteria to simulations
with vertical gravity. Also, with more than one particle species, we
expect numerical convergence with one particle per cell per particle
species, where different particle species have different stopping times.

Recently, \citet{Rein_etal10} addressed the validity of the super-particle
approximation in numerical simulations with particles. They emphasized
that it is essential to maintain the important timescales in the scaled
system (i.e. the simulation) to be equivalent to the timescales in the real
system, otherwise one cannot achieve numerical convergence. This is
unlikely to be relevant to the problem considered in this paper because
we have not added gravitational interaction nor particle collisions. The
only time scales to be fixed are the orbital period and the stopping time,
which are combined into the dimensionless stopping time $\tau_s$. Our
results imply that numerical convergence can still be achieved when
aerodynamic interaction between particle and gas is added, strengthen
conclusions of Rein et al.

\acknowledgments

We thank Anders Johansen and Andrew Youdin for helpful discussions
in the ``Dynamics of Discs and Planets" programme at the Newton
Institute, Cambridge University. We are grateful to Paul Bode and
Anatoly Spitkovsky for advises on the data structure of the particle
component, and Emmanuel Jacquet for his earlier development of the
particle module. We thank our referee for carefully reading the
manuscript and the helpful suggestions. This work is supported by NSF
grant AST-0908269. XNB acknowledges support from NASA Earth and
Space Science Fellowship.

\appendix

\section[]{Additional Linear Growth Rate Test of the Streaming Instability}\label{app:lin}

The linear and non-linear SI simulations in this paper adopt relatively
large particle stopping times $\tau_s\leq0.1$. In this appendix we provide
two additional linear growth test  problems to test code performance for
smaller $\tau_s$. The simulation set up is described section \ref{ssec:linSI},
the first test (``linC"), $\tau_s=0.01$, $\epsilon=2$, $K_x=K_z=1500$, the
analytical growth rate is $s=0.5981$. In the second test (``linD"),
$\tau_s=10^{-3}$, $\epsilon=2$, and $K_x=K_z=2000$, with growth rate
$s=0.3154$. Both modes are chosen to be close to the fastest growth mode.
However, these tests are more computationally costly because of the larger
$K$. The eigenvectors of the two modes are listed in Table
\ref{tab:eigensys}\footnote{We thank Andrew Youdin for providing the
eigenvectors.}. The table format is the same as Table 1 of YJ07.

\begin{figure*}
    \centering
    \includegraphics[width=145mm,height=65mm]{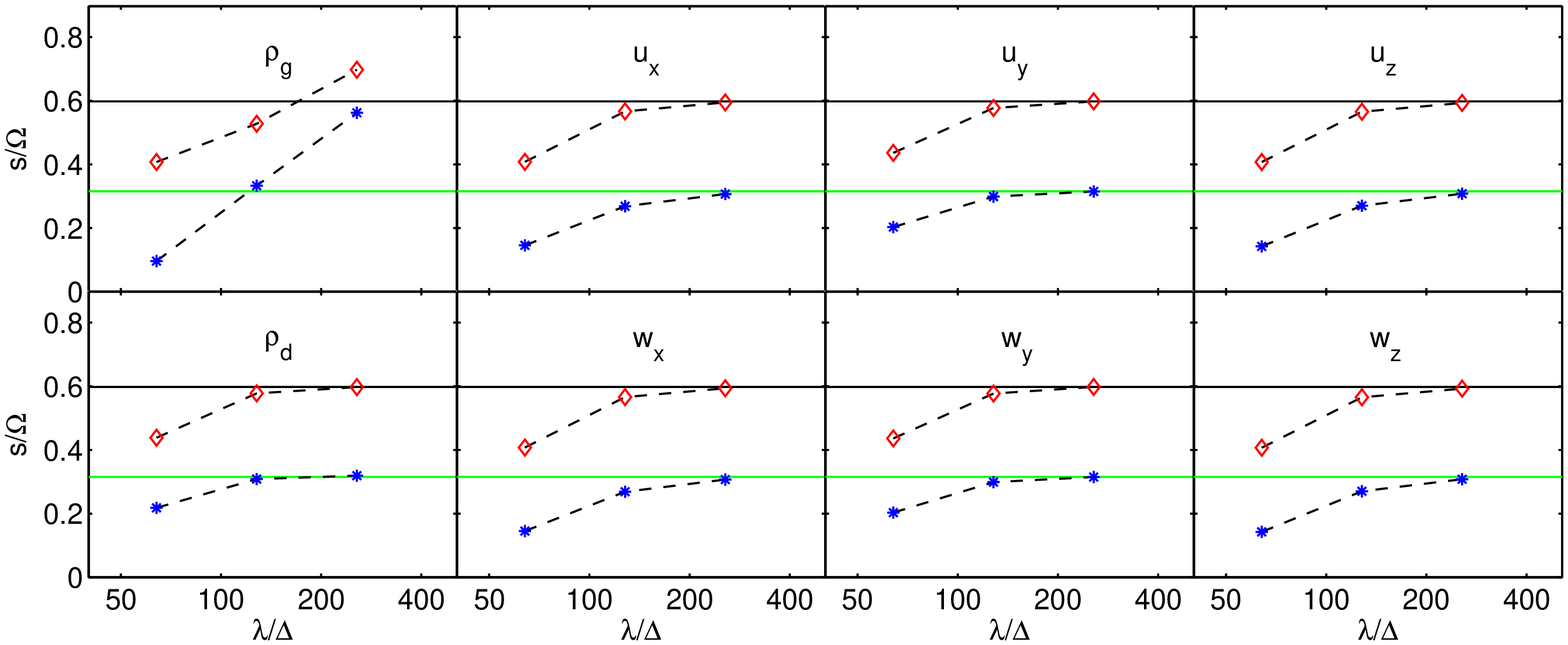}
     \caption{Measured growth rate of the streaming instability eigenmodes
  ``linC" (diamonds) and ``linD" (asterisks) using the semi-implicit integrator.
  Solid line marks the theoretical growth rate, and dashed lines with symbols
  label the measured growth rate as a function of grid resolution (number of
  cells per wavelength). We use CFL$=0.4$ in all the tests.}\label{fig:streaminglin2}
\end{figure*}

In Figure \ref{fig:streaminglin2} we show the numerical growth rate measured
as a function of grid resolution for the two tests ``linC" and ``linD". The CFL
number is fixed at 0.4 for these tests. Note that for smaller $\tau_s$, the
perturbations on the gas is extremely small, and capturing the correct growth
rate on the gas is very difficult. For other quantities, numerical convergence to
theoretical growth rate is clearly reached as shown in the figure, however, for
smaller $\tau_s$, higher resolution of about $128$ cells per wavelength is
needed for numerical convergence.

\begin{landscape}
\begin{table*}
\caption{Test Mode Eigensystems.}\label{tab:eigensys}
\begin{center}
\begin{tabular}{ccccccccccc}\hline\hline
 Test & ${\wt u}_x$ & ${\wt u}_y$ & ${\wt u}_z$ & ${\wt \rho}_g$ &
 ${\wt w}_x$ & ${\wt w}_y$ & ${\wt w}_z$ & $\omega$\\\hline
{\rm linC} $\tau_s=10^{-2},\ \epsilon=2$ &  $-0.1598751$ & $0.1164423$ & $0.1598751$
& $8.684872$e-8 & $-0.1567174$ & $0.1159782$ & $0.1590095$ & $0.1049236$\\
($K_x=K_z=1500$) & $+0.0079669i$ & $+0.0122377i$ & $-0.0079669i$
& $+5.350037$e-7$i$ & $+0.0028837i$ & $+0.0161145i$ & $-0.0024850i$ & $+0.5980690i$ \\
{\rm linD} $\tau_s=10^{-3},\ \epsilon=2$ &  $-0.1719650$ & $0.1918893$ & $0.1719650$
& $2.954631$e-7 & $-0.1715840$ & $0.1918542$ & $0.1719675$ & $0.3224884$\\
($K_x=K_z=2000$) & $+0.0740712i$ & $+0.0786519i$ & $-0.0740712i$
& $+1.141385$e-7$i$ & $+0.0740738i$ & $+0.0787371i$ & $-0.0739160i$ & $+0.3154373i$ \\
\hline\hline
\end{tabular}
\end{center}
Note: the table format is the same as Table 1 of YJ07.
\end{table*}
\end{landscape}


\label{lastpage}
\end{document}